\documentclass[12pt]{article}
\input epsf.sty
\usepackage{latexsym}

\newcommand{\CC}{{\cal C}}
\newcommand{\OO}{{\cal O}}

\newcommand{\LL}{{\cal L}}

\newcommand{\half}{{1\over 2}}

\newcommand{\be}{\begin{equation}}
\newcommand{\ee}{\end{equation}}
\newcommand{\ben}{\begin{eqnarray}\displaystyle}
\newcommand{\een}{\end{eqnarray}}
\newcommand{\refb}[1]{(\ref{#1})}

\newcommand{\sectiono}[1]{\section{#1}\setcounter{equation}{0}}

\begin{document}
{}~
\hfill\vbox{\hbox{hep-th/0207107}\hbox{CTP-MIT-3283}
}\break

\vskip 2.1cm

\centerline{\large \bf
Dynamics with  Infinitely Many Time Derivatives }

\vspace*{2.0ex}

\centerline{\large \bf
and Rolling Tachyons}

\vspace*{8.0ex}

\centerline{\large \rm Nicolas Moeller and Barton Zwiebach}

\vspace*{8.0ex}

\centerline{\large \it Center for Theoretical Physics}

\centerline{\large \it
Massachusetts Institute of Technology,}

\centerline{\large \it Cambridge,
MA 02139, USA}

\vspace*{2.0ex}
\centerline{E-mail: moeller, zwiebach@mitlns.mit.edu}

\vspace*{6.0ex}

\centerline{\bf Abstract}
\bigskip

Both  in string field theory and in p-adic string theory the 
equations of motion involve infinite number
of time derivatives. 
We argue that the initial value problem is qualitatively
different from that obtained in the limit of many time
derivatives in that the space of initial conditions becomes
strongly constrained. 
We calculate the energy-momentum tensor and study in
detail time dependent solutions representing tachyons rolling on the  p-adic
string theory potentials.  For even
potentials we find surprising small oscillations at the tachyon vacuum.
These are not conventional physical
states but rather anharmonic oscillations  with a nontrivial
frequency--amplitude relation.  When the
potentials are not even, small oscillatory solutions around the bottom must
grow in amplitude without a bound. Open string
field theory resembles this latter case, the tachyon
rolls to the bottom and  ever growing oscillations ensue. We discuss
the significance of these results for the issues of emerging
closed strings and tachyon matter.

\vfill \eject

\baselineskip=16pt

\tableofcontents

\sectiono{Introduction and Summary}\label{s0}
Some of the most intriguing and fascinating
aspects of string theory center around the
role of locality. More precisely, string theory
appears to be a consistent theory that is not
local in the sense of local quantum field theory.
Indeed, the covariant string field theory 
action governing the dynamics of the infinite
set of spacetime fields contains spacetime
derivatives of all orders. Not only we have
spatial derivatives of all orders, as in non-commutative
field theory, but we also have time derivatives
of all orders, and mixed derivatives of all orders.

The possible difficulties with theories having
a high  number of time derivatives are
familiar. They include possible violations of unitarity and causality,
the appearance of spurious solutions, complications setting up
an initial value problem, and difficulties quantizing
the theory and/or finding a stable Hamiltonian \cite{woodard}.  There is, however,
evidence that in the case of string theory the higher 
derivative structure does not threat the consistency of
the theory. The perturbative S-matrix of string theory is 
unitary.  Even though covariant string field theory
cannot be quantized using a Hamiltonian, path integral
quantization is possible and  elegantly done with the
use of the Batalin-Vilkovisky 
formalism \cite{bvpapers,thorn,9206084,siegel}. While
infinitely many spatial derivatives are possible in non-commutative
{\it field} theory, infinitely many mixed spatial and time derivatives
appear to require string theory for consistency
\cite{0005073,0103069}.
Finally, the higher derivative structure of string theory
is certainly very special -- when passing to the light-cone
gauge the theory becomes local in light-cone time. Indeed,
light-cone string field theory has only first order time 
derivatives \cite{kakukikkawa}.

Motivated by the recent work of Sen on tachyon 
matter \cite{Sen:2002nu,Sen:2002in,Sen:2002an}
we examine the rolling of the tachyon down its potential.
Tachyon matter is a classical open string theory solution where
at large times the tachyon $T(t)$ approaches the tachyon vacuum $T=\infty$
with
constant velocity $\dot T(t)$ . This solution represents a pressureless
gas and its possible cosmological relevance has been studied in 
\cite{tachMattCosm}.
The evidence for such solutions comes mostly from the conformal field
theory approach. Recently, related solutions have been obtained
from considerations of boundary string field 
theory \cite{Sugimoto:2002fp,Minahan:2002if}. These actions,
truncated
to include all powers of first derivatives of fields exhibit similar,
though not identical behavior.
The purpose of the present paper is to examine tachyon 
dynamics in the setup of
open string field theory \cite{witten}, 
where the equations of motion contain infinitely many
time derivatives. Most of our work, however, will concentrate in the
case of p-adic string theory \cite{Brekke}. This string theory is essentially
a theory of a scalar field with infinitely many spacetime derivatives.
Indeed many of the properties of the tachyon field of open string theory
are also exhibited by the p-adic 
string model \cite{Ghoshal:2000dd,Minahan:2001pd}. One has
a tachyon, and a potential with a locally stable minimum where the scalar field has
no conventional excitations. Lump-like solutions exist representing
p-adic D-branes of various dimensionalities. We will find here interesting
phenomena regarding tachyon solutions in p-adic string theory, and
we will also see that some solutions in string field theory are quite
analogous to p-adic string solutions. None of the solutions we find,
should be said, appears to represent tachyon matter.

\begin{figure}[!ht]
\leavevmode
\begin{center}
\epsfysize=6.0cm
\epsfbox{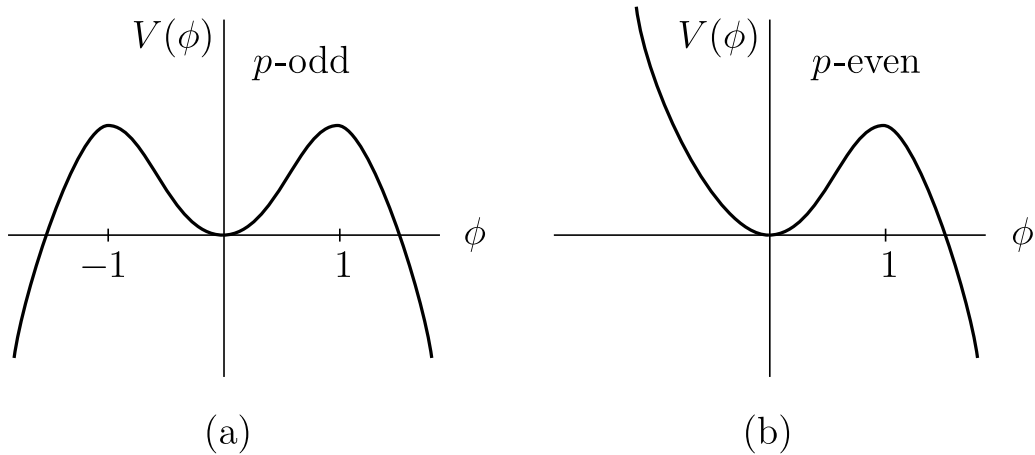}
\end{center}
\caption{The p-adic string potentials. For odd $p$ the potential
is even and has two unstable maxima. For even $p$ there is only
one unstable maximum.}
\label{mfig0}
\end{figure}

\medskip
In order to explain our results we briefly review the p-adic string model.
The model is defined by an action $S$ for a scalar field. Formulated with
an  arbitrary spacetime
dimensionality, the model has a parameter $p$, which is
initially taken to be a prime number. The action, however, makes sense for other
values of $p$, and even in the limit $p\to 1$, where it reduces to
the tachyon action of \cite{0008231} as noted in \cite{0009103}. 
In this paper we will take $p$ to be an integer that is greater
than or equal to two. The p-adic action is given 
by
\begin{equation}
\label{pmodel}
S = \int d^d x \LL
= {1\over g_p^2} \int d^d x \left[ -{1\over 2}
\phi\, p^{-\half\Box}\phi + {1\over p+1} \phi^{p+1} \right]\,,\,
\quad {1\over g_p^2} \equiv {1\over g^2} {p^2 \over p-1} .
\end{equation}
The infinite number of spacetime derivatives are
manifest in the differential operator $p^{-\half\Box}$.  Here, we have
\begin{equation}
\Box = - {\partial^2\over \partial t^2} + \nabla\cdot\nabla\,,
\end{equation}
and one defines
\begin{equation}
\label{expder}
p^{-\half\Box} = \exp \Bigl( -\half \ln p~ \Box\Bigr) =
\sum_{n=0}^\infty \Bigl( -\half \ln p\Bigr)^n{1\over n!}~  \Box^n\,.
\end{equation} 
The equation of
motion following from the action is rather simple looking:
\begin{equation}
 \label{eompadic}
p^{-\half\Box} \phi = \phi^p\, .
\end{equation}
For the case of time dependent but spatially homogeneous solutions --
the only kind we will study in this paper -- the equation is
\begin{equation}
 \label{teompadic}
p^{\half \,\partial_t^2}\, \phi = \phi^p\, .
\end{equation}
Finally, we note that the scalar potential is given by
\begin{equation}
V(\phi ) = {1\over g_p^2} \,\Bigl(\, {1\over 2}\,\phi^2 -{1\over p+1}\,
\phi^{p+1}\Bigr)\,.
\end{equation}
The cases odd $p$ and even $p$ are 
qualitatively different (see figure \ref{mfig0}),
although both have potentials unbounded from below. For odd $p$ the potentials
are even in $\phi$, while for even $p$ they fail to be even. In both
cases there is a local maximum at $\phi=1$, where the tachyon has
 $M^2 = - 2$, and a local minimum at $\phi=0$ where the field
seems naively to have no dynamics as its mass-squared goes to infinity.

\medskip
In order to study effectively these theories and their solutions
we need expressions for the components of the energy-momentum
tensor. For the case of a higher derivative theory
we extend the conventional Noether procedure and find an expression
for $T^\mu_{~\alpha}$ as an infinite series involving all numbers
of derivatives. These expressions are formally local and can
also be used for open string field theory. The energy of a solution
does involve infinite number of derivatives and it is conserved. 
Its evaluation can sometimes be done in closed form. It is
also determined unambiguously from the above Noether method. 
The Noether calculation of $T^i_i$, where from we must read the 
pressure, has improvement ambiguities and we
use the metric variation of the covariantized p-adic action to
calculate its value.  

\medskip
We also discuss the nature of the initial value problem
for equations such as \refb{teompadic} containing infinitely
many time derivatives.  While an equation of motion containing
up to $N$ derivatives requires $N$ independent initial conditions,
we show that in the case of infinite number of time
derivatives the infinite number of ``initial conditions" are
not  independent but are actually subject to 
(possibly) infinitely many constraints. It may even happen that
the space of consistent initial conditions is parametrized
by a finite number of parameters. If this were the case
we would have a surprising example of a theory
with infinitely many time derivatives whose solution space
is not altogether different to that following from 
equations with two time derivatives.   

\medskip
The first time dependent solution in p-adic string theory was
found by Brekke {\it et.al.}~\cite{Brekke}. The equation of motion
\refb{teompadic} was
studied by expressing the differential operator in the left hand side as a
convolution of
$\phi$ with a gaussian, as will be reviewed in section \ref{sectrev}.
The equation of motion then becomes a non-linear integral equation.
For the case of odd $p$ the authors found numerically a kink solution
of the integral equation. The solution interpolates in time between the
vacua $\phi =\pm 1$.  No solution was obtained for 
even $p$.

In this paper we study in detail time dependent solutions
of  p-adic string theory. We use a combination
of analytic and numerical methods. In particular we show
how to use the convolution approach to constrain solutions
and to rule out certain behaviors. Some of the numerical 
methods are based on considerations of 
\cite{Sen:2002nu,Sen:2002in}
and the recent work \cite{ashokenew}. While we
have no exact closed forms for any solution, we have 
analytic expressions valid in certain limits. As the solutions
are sometimes quite surprising we try to check them in two
ways. We verify they hold numerically both as solutions of
the nonlinear integral equation, and after evaluating several derivatives
of the solution, we confirm that they directly satisfy \refb{teompadic}
with the derivatives expanded as in \refb{expder}.  This not only
increases our confidence on the solutions, but it is of some theoretical
significance in that the exponentiated differential operator and
the convolution form do not agree when acting on certain kinds
of smooth functions.

After confirming the kink solution of \cite{Brekke} to high
accuracy and checking it does satisfy both versions of the
equation of motion, we turn to finding families of solutions
for odd $p$ potentials.  We find nontrivial field oscillations around
the tachyon vacuum $\phi=0$. These are rather surprising.
It is well known that the linearized equations of motion of
the tachyon around the vacuum admit no solutions.  This is
the p-adic version of the familiar statement that there are
no conventional open string excitations at the tachyon vacuum
\cite{senconj}.
The solutions we find exist because they solve the complete
nonlinear equations.  The oscillations are completely anharmonic
and the basic frequency goes to infinity as the amplitude of
the oscillation goes to zero.  The solutions are of the form
$\phi \simeq A (\cos (\omega t))^{1/p}$,  with $\omega^2
\sim -\ln (A)$ and provide an explicit
example of a non-standard open string excitation capable of carrying
energy at the tachyon vacuum.  These solutions are obtained by
a level analysis of the differential equation and were checked also
by the integral form.  There are two possible interpretations
for these oscillations. In the first one, they would represent
oscillating solutions with a nontrivial energy-momentum 
tensor which would be expected to radiate quite efficiently 
into closed string modes,
thus converting the tachyon energy into closed strings.
In the second interpretation these oscillations, suitably
quantized would be the closed string excitations themselves.
We will discuss both possibilities, but believe the former is
more likely to be correct.

\medskip
In the studies of tachyon matter the tachyon vacuum is at
infinity and the tachyon never reaches this point. Moreover, there is
no meaning to configurations beyond the tachyon vacuum.
On the other hand, in string field theory or in p-adic
theory, the tachyon vacuum is at a finite point in the field
configuration space and there are field values beyond the tachyon
vacuum.
In the solutions we obtained the field typically speeds by the
tachyon vacuum without difficulty. The case of even $p$, in particular
$p=2$, where the potential is cubic resembles open string
field theory. For even $p$  we have found no constant amplitude
oscillations around the tachyon vacuum (but we lack a proof that they 
cannot exist). For $p=2$ a rolling solution 
down the unstable maximum at $\phi=1$ is seen to 
zoom by the tachyon vacuum $\phi=0$ and after an excursion
to negative values  having potential energies higher than that
of the original unstable vacuum, ever growing
oscillations occur. In this solution the pressure does not
go to zero for large times.  The analog
for even
$p$ of the family of  odd $p$ bounded oscillations  is a
family of solutions that has ever growing oscillations
(ever growing oscillations  also seem possible  for odd $p$).

Last but not least we study the rolling problem in the 
full open string field theory
using the analytic continuation of the marginal deformation
solution of \cite{0007153} as advocated in \cite{Sen:2002nu}.
We find that the tachyon rolls down towards the vacuum, goes beyond
it by a large factor, and that
a pattern of growing oscillations appears to set it. This 
surprising result, in many ways, motivated our full analysis
of the p-adic string model. Indeed, the results summarized
above concerning the p-adic model indicate that rolling in
OSFT appear to be remarkably similar to rolling on the $p=2$ 
potential. This lends credence
to the idea that the OSFT solution can be trusted and
is not an artifact of some approximation scheme.  
On the other hand it also suggests
that this solution is not tachyon matter.  

Some concluding remarks and observations, as well
as a detailed list of open 
questions can be found in the conclusion section.

\sectiono{Convolution and Initial Value Problem}

In this section we begin by considering the convolution
form of the p-adic string equation of motion. In this
form, the equation is a nonlinear integral equation with
a gaussian kernel.  The convolution form and the differential
form of the equation of motion appear to be equivalent only
in the space of real analytic functions.

We then show that the convolution form of the equation
of motion is a useful tool that allows qualitative
analysis of solutions.  We use it to show that there cannot
exist solutions of p-adic string theory where the field
approaches  monotonically the tachyon vacuum for large times.
We also use it to find constraints on possible oscillatory
solutions and to rule out certain types of lump solutions.

Finally, we discuss the crucial issue of the initial
value problem in a theory with infinite number of 
derivatives.

\subsection{Convolution form of the p-adic equation of motion}
\label{sectrev}

In this subsection we discuss the convolution form
of the p-adic string dynamics obtained by \cite{Brekke}.
Let us first derive this form. To this end we recall the
differential form of the 
equation of motion (\ref{teompadic}):
\be
p^{{1\over2} \partial_t^2} \phi = \phi^p\,.
\label{EOMBox1}
\ee
As a first step we
assume a well defined  Fourier transform
$\hat{\phi}(k)$ of the p-adic field  $\phi(t)$:
$$
\hat{\phi}(k) = {1\over \sqrt{2 \pi}} \int_{-\infty}^\infty {e^{i k t} 
\,\phi(t) dt} \,,
$$
giving $\phi(t)$ by inverse Fourier transformation:
$$
\phi(t) = {1\over \sqrt{2 \pi}} \int_{-\infty}^\infty {e^{-i k t} 
\,\hat{\phi}(k) dk} \,.
$$
Then, the left hand side of the equation of motion \refb{EOMBox1} 
can be written as:
\ben
p^{{1\over2} \partial_t^2} \phi(t) &=& {1\over \sqrt{2 \pi}}
\int_{-\infty}^\infty {p^{-{1\over2} k^2} e^{-i k t} \hat{\phi}(k) dk}
\nonumber
\\
&=&
{1 \over \sqrt{2 \pi}} \int_{-\infty}^\infty dk \left( {1 \over \sqrt{2 \pi \ln p}}
\int_{-\infty}^\infty e^{-{{t'}^2 \over 2 \ln p}} e^{-i k (t - t')} dt' \right)
\hat{\phi}(k)
\nonumber
\\
&=& {1 \over \sqrt{2 \pi \ln p}} \int_{-\infty}^\infty
{e^{-{1\over 2 \ln p} (t-t')^2}
\phi(t') dt'}  \,.
\label{Boxtoconv}
\een
Defining the gaussian convolution
${\cal C}[\phi]$ 
\begin{equation}
\label{gconv}
{\cal C}[\phi](t) \equiv  {1 \over \sqrt{2 \pi \ln p}} 
\int_{-\infty}^\infty
{e^{-{1\over 2
\ln p} (t-t')^2}
\phi(t') dt'} \,,
\end{equation} 
the equation of motion can then be rewritten as
\be
\phi(t)^p = {\cal C}[\phi](t) 
\label{EOMconv} \,.
\ee
This is a nonlinear integral equation.
Note that the convolution operator is correctly normalized 
to represent the differential operator:  acting on
a constant it gives the constant back
\begin{equation}
{\cal C}[a] = a \,.
\end{equation}
Constant solutions $\phi(t) = \phi_0$ of the p-adic field
equation
\refb{EOMconv}  therefore arise when :
\begin{equation}
\phi_0^p = \phi_0\, \qquad  p>1.
\end{equation}
This gives $\phi_0=0$ and $\phi_0=1$ valid for all $p \geq 2$.
For odd $p$ we also have $\phi_0 = -1$. 
These solutions all correspond to the tachyon
sitting on a critical point of the p-adic string
potentials. 

Before we go on and solve either (\ref{EOMBox1}) or (\ref{EOMconv}), we
should note that there are smooth functions $\psi(t)$ ({\it i.e.~}continuous
functions with continuous derivatives of all orders) for which
the convolution does not represent the action of the differential
operator:
$$
p^{-{1\over2} \partial_t^2} \psi(t) \neq  {\cal C}[\psi](t)\,.
$$
A simple example is provided by the following smooth 
function with compact support:
\be
\psi(t) = \left\{ \begin{array}{c}
0, \quad |t| > b \\
1, \quad |t| < a\,. \end{array} \right. 
\ee
\begin{figure}[!ht]
\leavevmode
\begin{center}
\epsfxsize=12.0cm
\epsfbox{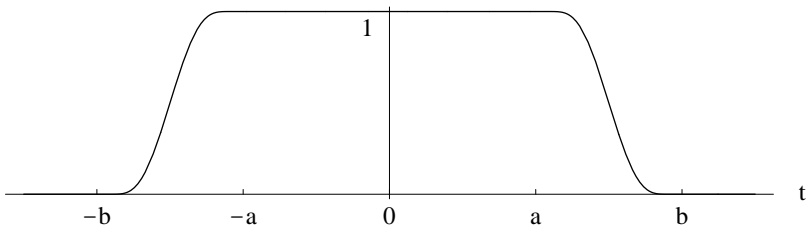}
\end{center}
\caption{A smooth function that is identically one for $|t|<a$, and
identically zero when $|t|>b$. For such a function the differential
form of the equation of motion does not coincide with the convolution
form.}
\label{smooth}
\end{figure}
where $a$ and $b$ are positive numbers such that $b>a$. When
$a \leq |t| \leq b$, $\psi$ interpolates smoothly between $0$ and $1$.
Since all the
derivatives of $\psi$ are zero at $t=0$, we have
$p^{-{1\over2} \Box} \psi(0) = \psi(0) = 1$; but on the other hand
${\cal C}[\psi](0) < 1$ since ${\cal C}[1] = 1$ and $\psi$ is strictly
smaller than one when $|t| > a$.
Therefore equation (\ref{Boxtoconv})  does not hold
for this smooth function. We expect, however, that (\ref{Boxtoconv})
holds for {\it real analytic} functions whenever the convolution
exists. We have checked many of the solutions
obtained in this paper and verified that they solve
both forms (\ref{EOMBox1}) and (\ref{EOMconv}) of the equations of motion.
This seems to mean that solutions of the p-adic string
theory lie on the space of real analytic functions.

\subsection{Convolution constraints on solutions}
\label{sectQA}

General features of a solution to the equation of motion
can be gleaned from the properties of the gaussian convolution.
We will illustrate this by proving three results that give
us insight into the questions we are trying to address.

The first result rules out the possibility that the
tachyon may roll monotonically down from $\phi=1$ reaching
the tachyon vacuum $\phi=0$. 

\medskip
\noindent
{\it  Claim 1: There is no solution $\phi(t)$ such that 
$\phi(t=-\infty) =1$ and $\phi(t=+\infty) =0$ with $\phi(t)$
decreasing monotonically in time.}

\begin{figure}[!ht]
\leavevmode
\begin{center}
\epsfysize=6.5cm
\epsfbox{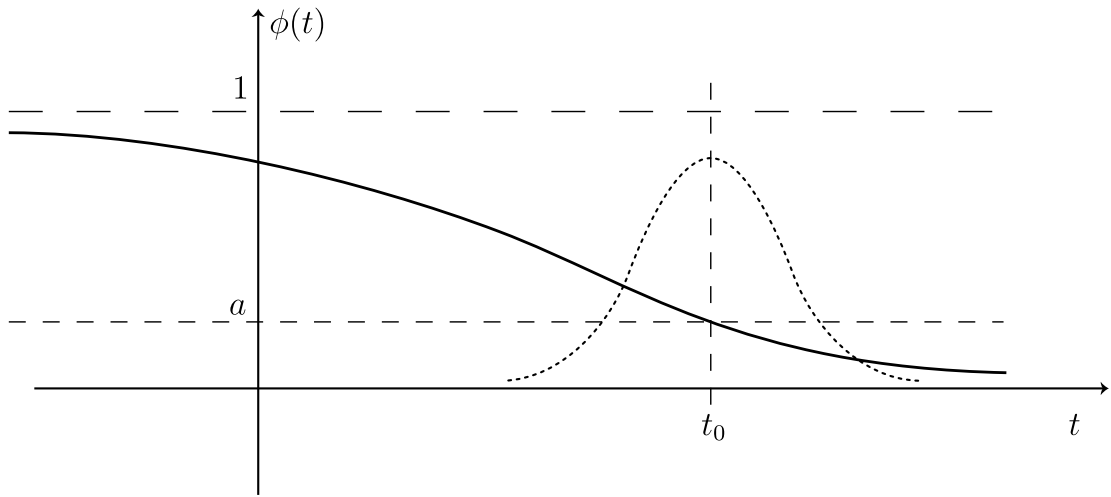}
\end{center}
\caption{A monotonically decreasing field configuration going from
the top of the potential all the way to the tachyon vacuum. Such field
configuration cannot satisfy the equation of motion.}
\label{mfig1}
\end{figure}

\medskip
\noindent
{\it Proof:}  Consider a monotonically decreasing function
$\phi(t)$ satisfying the above conditions, as illustrated
in figure \ref{mfig1}. Since the function approaches zero,
there is a time $t_0$ such that $\phi(t_0)=a$, with $a>0$ 
and sufficiently small such that 
\begin{equation}
\label{bineq}
a^p <  {a\over 2} \,.
\end{equation}
On the other hand
the equation of motion requires
\begin{equation}
\label{empro}
a^p = \CC [\phi] (t_0)\,.
\end{equation}
Consider now the auxiliary function $\psi(t)$ such that $\psi (t) = a$ for
$t< t_0$, and $\psi(t) = 0$ for $t>t_0$.  It is clear 
from the monotonicity property that $\phi(t) >
\psi(t)$ for all $t$. Moreover, $\psi$ has a very simple 
gaussian convolution. 
Using \refb{empro} we have  
\begin{equation}
a^p = \CC [\phi] (t_0) >\CC [\psi] (t_0) = {a\over 2}\,.
\end{equation}
This violates the inequality in \refb{bineq} and  proves the
claim that no solution exists satisfying the stated conditions.

\medskip
The second result concerns the possibility of solutions
bounded in time. These could be bounded oscillations or otherwise.
The ranges of variation are constrained by the following claim:

\medskip 
\noindent
{\it Claim 2: Consider a solution $\phi(t)$ and
 constants $a< b$ such that  
$a\leq \phi(t) \leq b$ for all times $t$. Moreover, assume that 
the values $a$, and $b$ are actually attained for some specific times.
Then $0<b<1$ and $a<0$. For odd $p$ we
must also have $a> -1$. }

\begin{figure}[!ht]
\leavevmode
\begin{center}
\epsfysize=6.5cm
\epsfbox{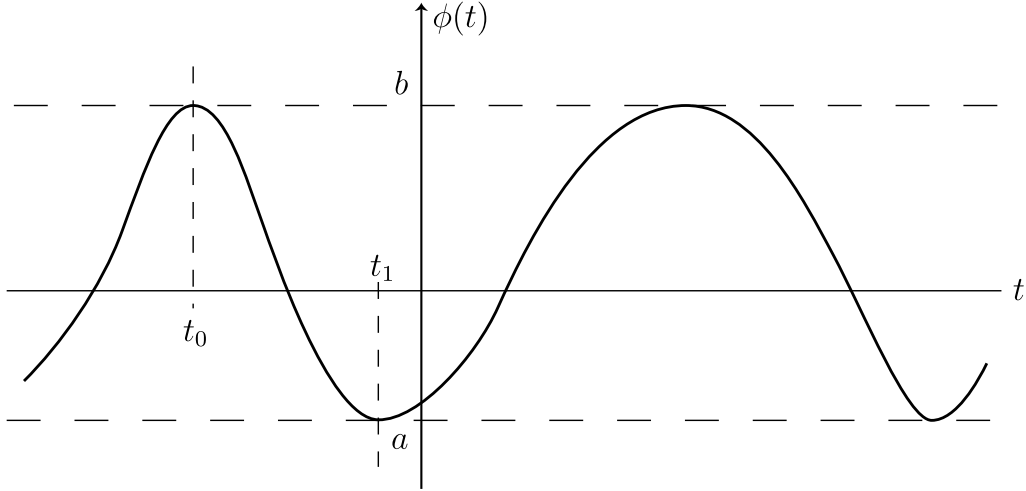}
\end{center}
\caption{A bounded field configuration attaining a maximum
value of $b$ and a minimum value of $a$. The field equation
gives constraints on the possible values of $a$ and $b$.}
\label{mfig2}
\end{figure}

\medskip
\noindent
{\it Proof:} Let $t_0$ denote a time when $\phi(t_0)= b$  and
$t_1$ denote a time when $\phi(t_1) = a$ (see Figure \ref{mfig2}).  From
the equation of motion we must have
\begin{equation}
b^p = \CC [\phi](t_0)  \,, \qquad  a^p = \CC [\phi](t_1) \,.
\end{equation}
Since $\phi(t_1) \leq \phi(t) \leq \phi(t_0)$  we have
\begin{equation}
b^p = \CC [\phi](t_0) < \CC[\phi(t_0)] = b  \,, \qquad  a^p = \CC
[\phi](t_1) > \CC
[\phi(t_1)] = a\,,
\end{equation}
giving us the inequalities
\begin{equation}
b^p  <  b  \,, \qquad  a^p  > a \,, \qquad b > a \,,
\end{equation}
where the last inequality is that from the definition of parameters.
We rewrite them as
\begin{equation}
b( b^{p-1} - 1)  <  0  \,, \qquad  a( a^{p-1} -1) > 0 \,, \qquad b > a \,.
\end{equation}
Note that $p-1$ is a positive integer.
Consider the $a$ inequality.  If both
factors are positive we must have $a>1$.  If both factors 
are negative there are two cases: (i) even $p$ requires $a<0$, 
and (ii) odd $p$ requires $-1 < a < 0$.  Consider now
the $b$ inequality.  If $b>0$  then $0< b< 1$.  
If $b<0$ there is no solution for even $p$, and for odd $p$
we see that actually $b<-1$.  The inequality $b>a$ allows
now selection of the final ranges.  Since $b<1$ in all cases, 
the possibility $a>1$ cannot be realized. Thus $a<0$ always.
If $p$ is even then $a<0$ and $0<b<1$.  If $p$ is odd then
$-1<a<0$.  This is not consistent with $b<-1$, and therefore
we get again $0<b<1$.  This is the statement in the claim.

\medskip
Our third result will concern the absence of certain
lump solutions for the case of even $p$. In order to establish
this result we first note that for any bounded function $\phi(t)$
the derivative of its convolution equals the convolution of 
its derivative:
\begin{equation}
\label{fder}
{d\CC[\phi]\over dt} (t) =  \CC \Bigl[ {d\phi\over dt}\Bigr] (t)\,.
\end{equation}
This result follows by noting that ${d\over dt}$ equals
to $\Bigl(-{d\over dt'} \Bigr)$ acting on
the gaussian appearing in the convolution \refb{gconv}. This derivative
can be integrated by parts without picking any boundary 
contributions and one can let it act on the function $\phi$ thus
obtaining the result. 

\begin{figure}[!ht]
\leavevmode
\begin{center}
\epsfysize=6.5cm
\epsfbox{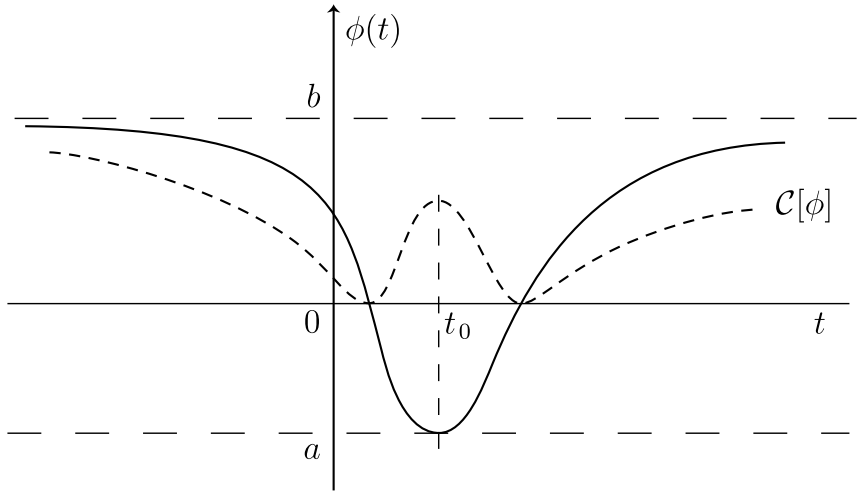}
\end{center}
\caption{A monotonic lump with endpoint values $b$ and 
unique minimum value of $a$. This kind of lump solution
cannot exist for even $p$. The dashed line represents the
gaussian convolution of the field configuration.}
\label{mfig3}
\end{figure}

\medskip
Let us now define a {\it monotonic lump} associated to
a bounded solution as in claim 2. This is a solution having
$\phi(-\infty) = \phi(\infty) = b$ and $\phi(t_0) =a$ for
some unique $t_0$.  In addition $\phi$ decreases monotonically
for $t<t_0$ and increases monotonically for $t>t_0$.  An example
of such $\phi$ is shown in figure \ref{mfig3}.

\medskip
\noindent
{\it Claim 3: There are no monotonic lump solutions
for even $p$.}

\medskip
\noindent
{\it Proof:} Because of claim 2, we have  $\phi(t_0) = a <0$.  Since
$\phi$ has a minimum
at $t_0$, the convolution $\CC[\phi](t)= \phi^p$ 
must have a maximum at $t_0$.  This requires
\begin{equation}
\label{mmax}
{d^2 \CC[\phi]\over dt^2} (t_0) < 0  \,.
\end{equation}
We differentiate \refb{fder} once more to obtain
\begin{equation}
{d^2 \CC[\phi]\over dt^2} (t_0) =  
{1 \over \sqrt{2 \pi \ln p}}\,
{1\over \ln p} \int_{-\infty}^\infty e^{-{1\over 2 \ln p}
 (t'-t_0)^2}
\Bigl[ (t'-t_0) {d\phi(t')\over dt'}\Bigr]  dt'\,.
\end{equation}
From the monotonicity condition we see that the factor in brackets
is positive both for $t'<t_0$ and for $t'>t_0$.  Therefore the above
integral is positive, in contradiction with \refb{mmax}. This establishes
claim 3.

\subsection{The fate of the  initial value problem}

In this subsection we will make some basic observations
about the initial value problem for the equation
\be
p^{{1\over2} \partial_t^2} \phi = \phi^p\,.
\label{EOMBox111}
\ee
In doing so, we view this as a differential
equation for time evolution. This equation has
an infinite number of derivatives and it is therefore
unclear how to define the initial value problem.

We will argue that an equation as the one above
is qualitatively very different from a differential
equation having a large but finite 
number of time derivatives. In solving
the initial value problem for a differential equation 
having time derivatives up to order $N$, one must 
specify $N$ initial conditions -- the values of the
field and those of its first $(N-1)$ derivatives
at an initial time, for
example. Typically, those values are unconstrained,
and solutions regular for some finite time exist
for any choice of initial conditions.

In an equation such as that in \refb{EOMBox111}
one would be led to believe that one can specify
an infinite number of independent initial conditions.  More
precisely, an infinite number minus one -- that is,
if we specify all derivatives of $\phi$ at zero time,
we could use the equation to read the value of $\phi$
at zero time.  But a puzzle arises -- if the function
and all of its derivatives are specified at zero
time, the natural assumption of analyticity would imply
that the function is known and can be reconstructed
from its Taylor expansion. It seems the differential equation
is not needed anymore!

We will see below that equation \refb{EOMBox111} 
actually imposes
an intricate set of conditions for the initial conditions!
We believe the analyticity assumption, and thus 
think that the evolution is determined fully by the
values of the function and all of its time derivatives
at the initial time -- nevertheless finding consistent
initial conditions is now {\it the problem}. 
 In the system with infinite number of time derivatives
{\it the problem is not that of evolving an initial value
configuration, but rather that of constructing a 
consistent initial value configuration}.

In an equation with up to $N$ derivatives, and with
$N$ initial conditions, taking further derivatives of
the equation typically does not yield constraints for the chosen
initial values. The resulting equations simply determine the higher
order derivatives at the initial time.  On the other hand 
for equation \refb{EOMBox111} taking arbitrary numbers of time 
derivatives yields new constraints that must be satisfied
by the infinite set of initial conditions.  This is
what makes the problem different.

To show that there are 
indeed many constraints implied in \refb{EOMBox111}
consider the following example.  Let the first derivative
of $\phi$ be non-vanishing at zero time, but all higher
derivatives vanish:
\begin{equation}
\label{icond}
\phi'(0) = a \not= 0, \quad \phi''(0)= \phi'''(0) = \ldots = 0\,.
\end{equation}
Consider now equation \refb{EOMBox111} for $p=2$. Consistent
with the above initial conditions we find
\be
p^{{1\over2} \partial_t^2} \phi (0) = \phi (0) = \phi^2(0)\,.
\label{testoem}
\ee
Consider now choosing $\phi(0) =1$ to satisfy this constraint.
Since we now have the field and all of its derivatives 
specified at zero time, the (analytic) field is thus
determined to be $\phi= 1 + at$, and this does not solve
equation \refb{EOMBox111}!  What went wrong?  We did not have
a consistent set of initial conditions. Indeed, taking two derivatives
of  equation \refb{EOMBox111} we have 
\be
p^{{1\over2} \partial_t^2} \partial_t^2 \phi (0) = 
2\phi'(0) \phi'(0) + 2\phi(0) \phi''(0) \,,
\label{testma}
\ee
which at zero time gives a new constraint on the initial
conditions
\be
0 =  2a^2  \,.
\label{testmax}
\ee
This constraint indicates that we should not have expected
a solution for nonzero $a$. It also shows that derivatives
of the equation of motion yields new constraints on the initial
values.

It does not seem altogether unreasonable to expect
that equation \refb{EOMBox111} together with all of its 
derivatives impose an infinite number of consistency 
conditions on the initial values. Could it be that  that consistent
initial values are parametrized by a finite number of parameters?
More exploration is necessary to answer this question.

\sectiono{Energy-Momentum in Higher Derivative Theories} \label{s2}

Both p-adic string theory and string field theory have
actions where fields appear with infinitely many derivatives,
both spatial and temporal.  It is also the case that
the action of p-adic  string theory has no explicit
dependence on the spacetime coordinates. 
 The same occurs for open string field
theory in many important backgrounds.  It therefore follows
that we expect to be able to define a conserved energy-momentum
tensor through a generalized Noether procedure. Since the Noether
procedure does not yield automatically a symmetric energy 
momentum tensor, we do not necessarily obtain the energy-momentum
tensor that acts as a source of the gravitational field.
This causes no complication for the case of the energy, which
is unambiguous.  But there are total derivative
ambiguities in identifying the pressure from the Noether 
construction of $(-T^i_i)$.  Therefore pressure is calculated
by doing a metric variation in the relativistic covariantization
of the p-adic action. We focus on the values of energy 
and pressure in the case
of time dependent but space independent backgrounds, and then
specialize further to the p-adic string model.

\subsection{Generalized Noether construction}

The situation in case is described by an action $S$ of the type

\begin{equation}
\label{actioninf}
S = \int d^Dx dt \,\,{\cal L}\, \Bigl(\phi,\,\, \partial_{\nu_1}\phi\,
,\,\,
\partial_{\nu_1}
\partial_{\nu_2}\phi \,,\,\, \partial_{\nu_1}
\partial_{\nu_2}\partial_{\nu_3}\phi\,, \,\, \cdots \Bigr)\,,
\end{equation}
where the dots denote the dependence on fields acted upon by
an ever increasing number of derivatives.  The symbol $\phi$
in the above action may represent a collection of fields,
but for notational simplicity
we will not include an extra index.  We emphasize that this
collection of fields can include, in addition to scalars, fields
of arbitrary tensor type.  This is indeed required for the results
to be useful in string field theory. All the results below are
trivially extended when $\phi$ is a collection of fields.

For a  Lagrangian without explicit coordinate dependence
the transformation
\begin{equation}
\label{ftrans}
\delta \phi = \epsilon^\alpha \partial_\alpha \phi,
\end{equation}
where $\epsilon^\alpha$ is a spacetime constant, gives
rise to a conserved current. Under this transformation
the Lagrangian varies into a total derivative
\begin{equation}
\delta \LL = \epsilon^\alpha \partial_\alpha \LL =  \partial_\alpha(
\epsilon^\alpha \LL) \,.
\end{equation}
If the lagrangian $\LL$ only contained first
order derivatives the stress tensor would be given as
\begin{equation}
\label{stens}
T^\mu_\alpha = - \delta^\mu_\alpha \, \LL
 +  {\partial \LL\over \partial \phi_\mu} \,\,\phi_\alpha \,.
\end{equation}
Here we are using the notation
\begin{equation}
\phi_\alpha \equiv \partial_\alpha \phi\,,
\end{equation}
or more generally, for any object $A$, we define
\begin{equation}
A_{\alpha_1\alpha_2\cdots \alpha_k} \equiv \partial_{\alpha_1}
\partial_{\alpha_2}\cdots \partial_{\alpha_k} A\,.
\end{equation}
The familiar result in \refb{stens}
holds for any field $\phi$ of any
tensor type\footnote{For fields other than
scalars the resulting energy-momentum tensor $T_{\mu\alpha}$ may not be
symmetric. }.
This is because under {\it constant}
space translations any tensor field transforms  as
indicated in \refb{ftrans}.
What we wish to have now
is the generalization of \refb{stens} for the case when
the action contains all possible numbers of derivatives.
Our result will be an expression for the stress tensor
involving all numbers of derivatives, and reducing to
the above if $\LL$ only depends on first derivatives.

We begin the derivation of the general formula
by writing the equation of motion following from the
action in \refb{actioninf}. One finds
\begin{equation}
\label{emotion}
0={\partial \LL\over \partial \phi} -
\Bigl( {\partial \LL\over \partial \phi_{\mu_1}} \Bigr)_{\mu_1}
+\Bigl( {\partial \LL\over \partial \phi_{\mu_1\mu_2}} \Bigr)_{\mu_1\mu_2}
-\,\cdots \,+ (-1)^k \Bigl( {\partial \LL\over \partial
\phi_{\mu_1\cdots\mu_k}}
\Bigr)_{\mu_1\cdots\mu_k} + \cdots \,\,.
\end{equation}
In here the  derivatives are defined as
\begin{equation}
{\partial \phi_{\mu_1\cdots \mu_k}
\over \partial \phi_{\nu_1\cdots \nu_k}} = \delta^{\nu_1}_{\mu_1} \cdots
\delta^{\nu_k}_{\mu_k} \,,
\end{equation}
with no additional symmetrizations. Thus while $\phi_{\mu_1\cdots \mu_k}$
is by definition symmetric in all indices, ${\partial\over
\partial\phi_{\nu_1\cdots \nu_k}}$ is not.  It is
also useful to note that
\begin{equation}
\label{deraction}
\partial_\alpha \LL = {\partial \LL\over \partial \phi} \, \phi_\alpha
+ {\partial \LL\over \partial \phi_{\nu_1}} \, \phi_{\alpha\nu_1}
+ {\partial \LL\over \partial \phi_{\nu_1\nu_2}} \,
\phi_{\alpha\nu_1\nu_2}
+ \cdots + {\partial \LL\over \partial \phi_{\nu_1\cdots\nu_k}} \,
\phi_{\alpha\nu_1\cdots\nu_k} + \cdots \,\,.
\end{equation}
This equation encodes  the lack of explicit coordinate
dependence in the lagrangian.

At this stage it is simplest to state the result for the
energy-momentum tensor and then confirm it is conserved.
We find
\begin{eqnarray}
\label{ttensor}
T^\mu_\alpha =&& - \delta^\mu_\alpha \, \LL \cr\cr
&& +  {\partial \LL\over \partial \phi_\mu} \,\,\phi_\alpha \cr\cr
&&-\Bigl\{ \Bigl( {\partial \LL\over \partial \phi_{\nu_1\mu} }
\Bigr)_{\nu_1}
\,\phi_\alpha -\Bigl( {\partial \LL\over \partial \phi_{\mu\nu_1} } \Bigr)
\,\phi_{\alpha\nu_1} \,\Bigr\}
\cr\cr
&&+\Bigl\{\, \Bigl( {\partial \LL\over \partial \phi_{\nu_1\nu_2\mu} }
\Bigr)_{\nu_1\nu_2}
\,\phi_\alpha -\Bigl( {\partial \LL\over \partial \phi_{\nu_1\mu\nu_2} }
\Bigr)_{\nu_1}
\,\phi_{\alpha\nu_2}  +
\Bigl( {\partial \LL\over \partial \phi_{\mu\nu_1\nu_2} } \Bigr)
\,\phi_{\alpha\nu_1\nu_2} \Bigr\} \cr
&&\qquad \vdots \cr
&& + (-1)^k \Bigl\{ \,
\Bigl( {\partial \LL\over \partial \phi_{\nu_1\nu_2\cdots \nu_k\mu} }
\Bigr)_{\nu_1\nu_2
\cdots \nu_k}
\,\phi_\alpha
-\Bigl( {\partial \LL\over \partial \phi_{\nu_1\nu_2\cdots \mu\nu_k} }
\Bigr)_{\nu_1\nu_2
\cdots \nu_{k-1}}
\,\phi_{\alpha\nu_k} \cr\cr
&& \qquad\qquad \cdots +(-1)^k
\Bigl( {\partial \LL\over \partial \phi_{\mu\nu_1\nu_2\cdots \nu_k} }
\Bigr)
\,\phi_{\alpha\nu_1\nu_2
\cdots \nu_k} \Bigr\}  \cr
&& \quad \vdots
\end{eqnarray}
We must now verify that
\begin{equation}
\partial_\mu T^\mu_\alpha =0 \,,
\end{equation}
when we use the equation of motion \refb{emotion} and equation
\refb{deraction}.
The verification is straightforward once we note that each group of terms
in between braces has a simple derivative. 
When taking a $\partial_\mu$ on
any such group the only contributions are from: $\partial_\mu$ acting on
the
$\LL$ derivative in the first term, and, from $\partial_\mu$
acting on the field derivative in the last term. All other derivatives
cancel each other due to the alternating signs in the set of terms.
As a result we find
\begin{eqnarray}
\partial_\mu T^\mu_\alpha =&& - \partial_\alpha \, \LL \cr\cr
&& + \Bigl( {\partial \LL\over \partial \phi_\mu}\Bigl)_\mu
\,\,\phi_\alpha
+ \Bigl( {\partial \LL\over \partial \phi_\mu}\Bigl) \,\,\phi_{\alpha\mu}
\cr\cr
&&+\Bigl\{- \Bigl( {\partial \LL\over \partial \phi_{\nu_1\mu} }
\Bigr)_{\nu_1\mu}
\,\phi_\alpha +\Bigl( {\partial \LL\over \partial \phi_{\mu\nu_1} } \Bigr)
\,\phi_{\alpha\mu\nu_1} \,\Bigr\}
\cr\cr
&&+\Bigl\{\, \Bigl( {\partial \LL\over \partial \phi_{\nu_1\nu_2\mu} }
\Bigr)_{\nu_1\nu_2\mu}
\,\phi_\alpha  +
\Bigl( {\partial \LL\over \partial \phi_{\mu\nu_1\nu_2} } \Bigr)
\,\phi_{\alpha\mu\nu_1\nu_2} \Bigr\} \cr
&&\qquad \vdots \cr
&& +  \Bigl\{ \,(-1)^k
\Bigl( {\partial \LL\over \partial \phi_{\nu_1\nu_2\cdots \nu_k\mu} }
\Bigr)_{\nu_1\nu_2
\cdots \nu_k\mu}
\,\phi_\alpha  +
\Bigl( {\partial \LL\over \partial \phi_{\mu\nu_1\nu_2\cdots \nu_k} }
\Bigr)
\,\phi_{\alpha\mu\nu_1\nu_2
\cdots \nu_k} \Bigr\}  \cr
&& \quad \vdots
\end{eqnarray}
Using the equation of motion \refb{emotion} one finds that the
terms multiplying $\phi_\alpha$ add up to
${\partial \LL\over \partial \phi}$. Using \refb{deraction} one then
sees that indeed $\partial_\mu T^\mu_\alpha =0$.

As written, the expression for the energy-momentum
tensor can be used directly to obtain its form in p-adic
string theory. As mentioned before,  this
result can be readily extended to a collection of fields $\phi^i$. In this
case
one simply sums over the various fields  by replacing in \refb{ttensor}
the
terms
\begin{equation}
{\partial \LL\over \partial \phi_{\ldots} }
\,\phi_{\ldots}  \to {\partial \LL\over \partial \phi_{\ldots}^i }
\,\phi_{\ldots}^i
\end{equation}
With this replacement we can use \refb{ttensor} to find
the energy-momentum tensor associated to a solution in
open string field theory.

\subsection{Energy in time dependent solutions}

The focus in this paper is on solutions
that have time dependence but no spatial dependence.
For these class of solutions the expression \refb{ttensor}
for the energy momentum tensor simplifies considerably
and all the tensor components can only have time dependence.
From the conservation law, and with 
$i,j,$ denoting spatial indices, 
\be
0= \partial_\mu T^\mu_0 = \partial_0 T^0_0+ \partial_i T^i_0 = 
\partial_0 T^0_0\,.
\ee
Since the energy $T^0_0 (t)$ is conserved we do not
have ambiguities in its construction -- an additive
constant ambiguity is fixed by the condition that
when the field does not vary in time, the energy must
coincide with the potential energy.  

For solutions
with no spatial dependence we also have that
\begin{equation}
\phi_{\mu_1 \cdots\, i\, \cdots \mu_k} =0 \,,
\end{equation}
since all space derivatives must vanish.
Therefore, the terms involving field derivatives 
in \refb{ttensor} vanish
when calculating $T^0_i$ and $T^i_j$.  We find 
\begin{equation}
T^0_i = 0 \,, \qquad T^i_j = -\delta^i_j \LL \,.
\end{equation}
Note that the conservation law $\partial_\mu T^\mu_i =0$ 
gives $\partial_0 T^0_i+ \partial_j T^j_i =0$
or equivalently $\partial_j T^j_i =0$. This
conservation does not rule out time dependent improvement
terms in $T^j_i$. Therefore the identification of the pressure
as $p(t) = - T_{11} = - T_{22} = \cdots = + \LL (t)$
is ambiguous. This is also clear from the fact that
it is set to equal the Lagrangian, a quantity for which
total time derivatives are ambiguous. We will give a
computation of the pressure using the methods of general
relativity in the following subsection.

Let us now calculate the energy density
$T^0_0$. Denoting, for any quantity $A(t)$
\begin{equation}
A_n \equiv  {\partial^n \over \partial t^n } A\,,
\end{equation}
we find from \refb{ttensor} that the $k^{th}$ term
in braces becomes
\begin{equation}
 (-1)^k \Bigl\{ \,
\Bigl( {\partial \LL\over \partial \phi_{k+1} } \Bigr)_{k}
\,\phi_1
-\Bigl( {\partial \LL\over \partial \phi_{k+1} } \Bigr)_{k-1}
\,\phi_{2}  +\cdots +(-1)^k
\Bigl( {\partial \LL\over \partial \phi_{k+1} } \Bigr)_0
\,\phi_{k+1} \Bigr\}\,,
\end{equation}
and can be rewritten as
\begin{equation}
\sum_{m=0}^k (-1)^m
\Bigl( {\partial \LL\over \partial \phi_{k+1} } \Bigr)_{m}
\,\phi_{k+1-m} \,.
\end{equation}
We then recognize that the complete expression for the
energy is simply
\begin{equation}
E = T^0_0 = - \LL + \sum_{k=0}^\infty\sum_{m=0}^k (-1)^m
\Bigl( {\partial \LL\over \partial \phi_{k+1} } \Bigr)_{m}
\,\phi_{k+1-m} \,.
\end{equation}
Shifting the sum over $k$ by one unit we write
\begin{equation}
\label{energyf}
E(t) = - \LL \,\,+ \,\,\sum_{k=1}^\infty\sum_{m=0}^{k-1} (-1)^m
\Bigl( {\partial \LL\over \partial \phi_k } \Bigr)_{m}
\,\phi_{k-m} \,.
\end{equation}
This is the result we needed.  It provides an expansion
for the energy in terms of time derivatives of the field.
If the lagrangian depends only on finite number of derivatives
then the expansion terminates for some finite value of $k$.
If the lagrangian has infinite number of time derivatives
then the expression for the energy is an infinite series.
That series may or may not be possible to sum in closed
form. Nevertheless the expression for the energy is formally
local -- it gives the value of $E(t)$ in terms of derivatives
all of which are evaluated at $t$.

\medskip
We will now evaluate the above  expression for the energy 
 in the case of the p-adic string theory \refb{pmodel}.
For any solution $p^{-\half\Box} \phi = \phi^p$ of the
equation of motion the 
lagrangian density evaluated at the solution becomes
\begin{equation}
\LL = {1\over g_p^2}  \, {1\over 2} \, {1-p\over 1+p} \, \phi^{p+1}\,.
\end{equation}
In addition, the derivative  ${\partial \LL\over \partial \phi_k }$
only exists for even $k$, and reads:
\begin{equation}
{\partial \LL\over \partial \phi_{2\ell} } = -{1\over 2g_p^2} \phi \,\Bigl(
{1\over 2} \ln p\Bigr)^\ell {1\over \ell!} \,.
\end{equation}
With this result, the expression for the energy in \refb{energyf} becomes
\begin{equation}
E(t) = - \LL \,\,+ \,\,{1\over g_p^2}\sum_{\ell=1}^\infty\sum_{m=0}^{2\ell-1}
-{1\over 2} \,\Bigl( {1\over 2}
\ln p\Bigr)^\ell {1\over \ell!}(-1)^m \phi_m \,
\,\phi_{2\ell-m}\,,
\end{equation}
or, after minor rearrangements, 
\begin{equation}
E(t) = - \LL \,\,- \,\,{1\over 2g_p^2}\, \sum_{\ell=1}^\infty
\Bigl( {1\over 2}
\ln p\Bigr)^\ell {1\over \ell!}\sum_{m=0}^{2\ell-1}
\,(-1)^m \phi_m
\,\phi_{2\ell-m} \,.
\label{Epadic}
\end{equation}
This formula will be used to confirm the correctness
of certain solutions of p-adic string theory by testing
energy conservation.  We will also use it in section 
\ref{sectAO} to compute the energy of oscillatory solutions.

\subsection{Calculation of the pressure}

In this section we use the 
unambiguous definition of the energy-momentum
tensor from general relativity.  We make the
p-adic model action invariant under coordinate
transformations by including a spacetime metric
$g_{\alpha\beta}$ and calculate the energy-momentum
tensor as 
\be
\label{emtnesor}
T_{\alpha\beta} = {2\over \sqrt{-g}} 
{\delta S\over \delta g^{\alpha\beta}}\,.
\ee
The action of the p-adic model, with derivatives
expanded and covariantized reads
\begin{eqnarray}
\label{pmodelcov}
S =  {1\over g_p^2} \int d^d x\sqrt{-g}\,  \Bigl( -{1\over 2}
\phi^2 + {1\over p+1} \phi^{p+1} \Bigr)
- {1\over 2g_p^2} \sum_{\ell=1}^\infty \Bigl(-{1\over 2} \ln p\Bigr)^\ell
{1 \over \ell!} \int d^d x \sqrt{-g} \phi \, \Box^\ell \, \phi \,.
\end{eqnarray}
Here the box operators are the covariant ones, and since all
of them act on scalars they can be all written as 
\be
\Box  \, \phi = {1\over \sqrt{-g}} \partial_\mu 
( \sqrt{-g}g^{\mu\nu}\partial_\nu \, \phi  ) \,.
\ee
The  terms that are delicate to vary 
involve repeated action of box operators:
\begin{eqnarray}
\label{seriesd}
\int dx \sqrt{-g} \phi\, \Box^\ell \,\phi &=& \int dx\,\, \phi 
\,\, \partial_{\mu_1} 
 \sqrt{-g}g^{\mu_1\nu_1}\partial_{\nu_1} \,\, 
\,{1\over \sqrt{-g}} \partial_{\mu_2} 
 \sqrt{-g}g^{\mu_2\nu_2}\partial_{\nu_2} \,\,\cdots \nonumber\\ 
&&  \cdots {1\over \sqrt{-g}} \partial_{\mu_\ell} 
\sqrt{-g}g^{\mu_\ell\nu_\ell}\partial_{\nu_\ell} \, \phi \,,  
\end{eqnarray}
where all derivatives act to the right.  Since we are interested
in pressure we will only consider the computation of the components
of the energy-momentum tensor $T_{\alpha\beta}$ having spatial indices. Since 
the solutions we focus on are only time dependent, this means that
we need not vary the metric components $g^{\mu_i\nu_i}$ appearing
above  for these
would yield contributions to the pressure that have explicit 
spatial derivatives. It
therefore suffices to vary the factors of $\sqrt{-g}$, 
which is done with the help of 
\be
\delta \sqrt{-g} = -{1\over 2} \sqrt{-g} \,\,g_{\alpha\beta} \,\delta g^{\alpha\beta}\,,
\quad
\delta {1\over \sqrt{-g} }= {1\over 2} {1\over \sqrt{-g}}\,\,g_{\alpha\beta} 
\,\delta
g^{\alpha\beta}\,.
\ee
When we vary a particular $\sqrt{-g}$ in \refb{seriesd}
we must integrate by parts all the derivatives to the left
of the variation. There are $(2\ell-1)$ such variations in \refb{seriesd},
each one giving a different splitting of the derivatives between
the two fields.  All in all we find (with $i$ not summed) that the 
pressure is given by
 \begin{eqnarray}
\label{pform}
p(t) = - T^i_i =  {1\over g_p^2} \Bigl( - {1\over 2} \phi^2 + {1\over p+1} \phi^{p+1}
\Bigr)  + {1\over 2g_p^2} \sum_{\ell=1}^\infty
\,\Bigl( {1\over 2} \ln p\Bigr)^\ell {1\over \ell!}
\sum_{m=1}^{2\ell-1} \phi_m \phi_{2\ell-m} \,.
\end{eqnarray}
This pressure formula will be used to analyze rolling
solutions.

\sectiono{Rolling Down the Unstable p-adic Vacuum}

In this section we construct solutions representing rolling
of the tachyon from the unstable vacuum. That is, we want solutions
where in the infinite past the tachyon approaches the unstable 
vacuum.  We will obtain a solution for the case of 
odd $p$ (even potentials) and another very different solution for the case
of even $p$.  

It was already pointed out in \cite{Brekke} that 
(\ref{EOMconv}) has a
kink and an anti-kink solution when $p$ is odd. These are the
solutions we are interested in for they  start on one maximum at $t
\rightarrow -\infty$. As it turns out they  end up on the other maximum
at $t\rightarrow +\infty$. We will construct such solutions
to high accuracy, and verify that they not only solve the convolution
form of the field equation, but also the differential form.

The case of even $p$ is much more surprising.  Here one could
have expected a lump solution, but as we saw in section \ref{sectQA}
these are not expected to exist. Essentially the difficulties
arise because even though the field can go into negative values
the convolution, that must equal an even power of the field,
cannot. What we
find (for $p=2$) is rolling down, an overshooting and then ever-growing 
oscillations. This is the result of a calculation using an
ansatz for the solution in the form of a sum of exponentials. The result
is so surprising that we check it in two ways.  One by verifying
that the solution thus constructed satisfies the convolution
form of the field equation, and second, by evaluating the energy
and checking that it is conserved for the indicated motion. We also
calculate the pressure and find that it does not go to zero.
We will see in section \ref{sectOSFT} that this behavior seems
to occur also in OSFT.

\subsection{Kink solution for odd $p$}
\label{sectksop}

Even though the analytical form of the kinks are not known, it is
relatively easy to construct them numerically. For example, one can
use the iterative procedure where a given field configuration
$\phi(t)$ is used to calculate the function $\CC[\phi]$. The
field equation then implies that $(\CC[\phi])^{1/p}$ gives
a new (and possibly improved) approximation to $\phi(t)$. Thus
we hope for a situation where the iteration 
\be
\phi(t) \longrightarrow 
\left({1 \over \sqrt{2 \pi \ln p}} \int{e^{-{1\over 2 \ln p} (t-t')^2}
\phi(t') dt'}\right)^{1/p} \,,
\label{iteration}
\ee
converges to some definite answer.  Note that for odd $p$ there
is no complication taking the root -- positive numbers are taken
to have positive roots, and negative numbers are taken to have
negative roots.

\begin{figure}[!ht]
\leavevmode
\begin{center}
\epsfysize=7.5cm
\epsfbox{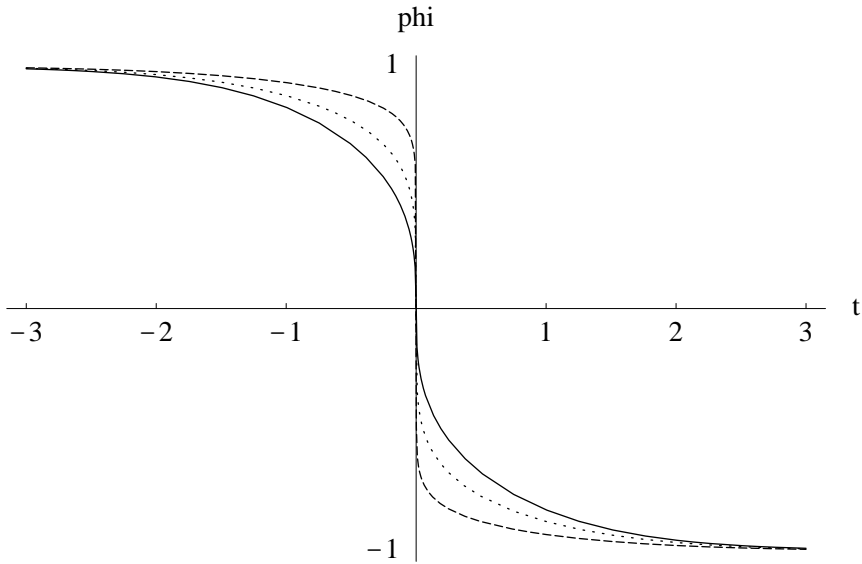}
\end{center}
\caption{Kink solutions for $p=3$ (solid line), $p=5$ (dotted line),
and $p=11$ (dashed line). The horizontal axis is time, and the value of
the field $\phi(t)$ interpolates between the unstable vacuum $\phi=1$
in the far past and the unstable vacuum $\phi=-1$ in the far future.}
\label{kinks}
\end{figure}

Starting
with the step function
\be
\phi(t) = \left\{ \begin{array}{rc} +1, & \quad t<0\,, \\
 0, & \quad t=0\,, \\
-1, & \quad t>0\,, 
\end{array} \right. 
\ee
$\phi(t)$ will converge to the kink solution when we iterate
(\ref{iteration}). Figure \ref{kinks} shows the kink
solutions
for $p=3$, $p=5$ and $p=11$. As can be seen from (\ref{EOMconv})
(and as already noted in \cite{Brekke}), near $\phi=0$ the field
behaves like $\phi(t) \sim t^{1/p}$. Also when $p \rightarrow \infty$,
the kink tends to a step function. Note that this means the field
zooms by the tachyon vacuum crossing it with infinite velocity and
spending the least possible amount of time in its vicinity. 

Our numerical kink solutions have been constructed from (\ref{EOMconv}).
As mentioned before, it is important to check that they also satisfy
equation (\ref{EOMBox1}).   From
equation \refb{expder} we have that 
\begin{equation}
\label{expderp}
p^{-\half\Box} = \exp \Bigl( \half \ln p~ \partial_t^2\Bigr) =
\sum_{m=0}^\infty \Bigl( { \ln p\over 2} \Bigr)^m{1\over m!}~ 
\partial_t^{2m}\,.
\end{equation} 
We are thus led to define an approximate version $D^{(2n)}$ of this
differential operator where we include time derivatives of order less than
or equal to $2n$: 
\be
D^{(2n)}\phi(t_0) \equiv \sum_{m=0}^n{\left({\ln p \over 2}\right)^m
{1 \over m!} \left( \partial_t \right)^{2m}} \, \phi(t_0) \,.
\ee
In Table \ref{DerTable} we perform a check of the solutions
obtained for $p=3$. We choose two values of the field 
$\phi = 0.999, ~0.99$ to do our check.  For these
values we compare
$\phi^3$ to 
$D^{(2n)} \phi$, where the derivatives are evaluated
using the numerical solution for the $p=3$ kink. 
If $\phi(t_0)$ satisfies
(\ref{EOMBox1}), then
$D^{(2n)}\phi(t_0)$ should converge to $\phi^3(t_0)$ when $n \rightarrow
\infty$.
In the first case, only
eight derivatives can be evaluated accurately; in the second
case, we can trust only six derivatives. In both cases, this suffices
to see a relatively good agreement between $D^{(2n)}\phi(t_0)$ and
$\phi^3(t_0)$. We thus conclude that the kink obtained by
the convolution form of the equation of motion satisfies 
also the differential form.

\begin{table}
\begin{center}
\begin{tabular}{|c|c||c|c|c|c||c|}  \hline
$t_0$ & $\phi(t_0)$  & $D^{(2)}\phi(t_0)$  & $D^{(4)}\phi(t_0)$  &
$D^{(6)}\phi(t_0)$ & $D^{(8)}\phi(t_0)$ & $\phi^3(t_0)$
\\  \hline \hline
$-4.67086$  & 0.999  & 0.997905  & 0.997309 & 0.997067 & 0.996974 &
0.997002
\\  \hline \hline
$-3.04298$ & 0.99  & 0.978765  & 0.972588  & 0.971867  &  & 0.970299
\\ \hline
\end{tabular}
\end{center}
\caption{Checking eqn.~(\ref{EOMBox1}) on the numerical kink
solution ($p=3$) computed from (\ref{EOMconv}).}
\label{DerTable}
\end{table}

If we try to repeat the same check for smaller values of the field,
we see that high derivatives have large absolute values, and it
would then be necessary to sum a large number of them to observe
convergence. Finer numerical methods would be needed to evaluate enough
derivatives. It will become clear in section \ref{sectAO}
that the kink is
the zero-frequency limit of solutions describing anharmonic
oscillations; and this will give us further evidence that the
differential form of the equation
is indeed satisfied for the kink.

\subsection{Rolling tachyons for p-adic strings with even $p$}
\label{sectMRTEP}

The case of even $p$ is rather different from the case of
odd $p$. When $p$ is even the potential has only one maximum 
($\phi=1$) and we cannot have a kink solution. We could expect
that the kink gets replaced by a lump-like solution, but
as we showed in section \ref{sectQA}, a typical lump solution
does not exist. We will see that if the tachyon rolls down
the unstable maximum towards the tachyon vacuum it overshoots
it and then ever growing oscillations begin.  We will first
argue qualitatively why such behavior occurs. Then we will
calculate the field evolution using a suitable expansion.
Finally we will test the solution in two ways.

\subsubsection{Qualitative discussion of rolling solution} 

\begin{figure}[!ht]
\leavevmode
\begin{center}
\epsfysize=6.5cm
\epsfbox{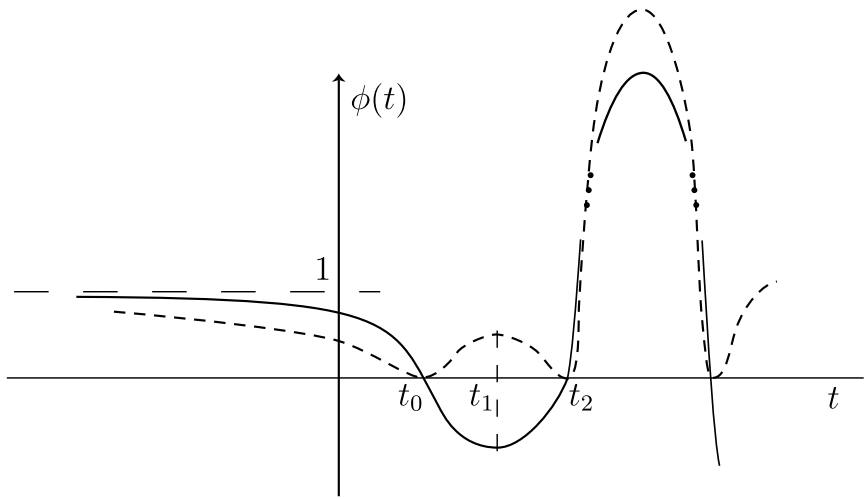}
\end{center}
\caption{A rolling solution for even $p$ and a consistency
analysis based on the value of the  convolution $\CC[\phi]$
shown in dashed lines.}
\label{mfig4}
\end{figure}

We now imagine the tachyon starts to roll down from
$\phi = 1$ in the infinite past towards the tachyon
vacuum $\phi=0$.  Because of claim 1 of section \ref{sectQA}
we know that the field cannot go asymptotically to
the tachyon vacuum for large time.  Thus we expect the
field to go past $\phi=0$ into negative values, as illustrated
in figure \ref{mfig4} where at some time $t_0$ we have 
 $\phi(t_0)=0$.  As soon as the
field turns negative it becomes delicate satisfying the equation
of motion, since the convolution must remain positive. This
would be very hard to achieve if the field remains negative
for a long time.  We can therefore expect the field to
reach a minimum at some time $t_1$  and then go back to
zero value at some time $t_2$.

In  figure \ref{mfig4} a dashed line shows the convolution,
which must equal $\phi^p$.
The convolution at $t_0$ must be zero, and this is possible
since the field is substantially negative to the right of $t_0$. 
But right after $t_0$ the convolution must increase. How can
that happen when the gaussian peak is moving towards values
where $\phi$ is turning more negative ?  This can happen
if after $t_2$ the field $\phi(t)$ is becoming positive very quickly.
It must become so large and positive that even the fast-decreasing 
gaussian tail manages to pick up a significant contribution. 

Now consider the convolution at $t_1$. It can be positive because
of the large positive values of the field beyond $t_2$. But this
convolution must  start decreasing to the right of 
$t_1$. It must do so even though the gaussian peak is moving
towards values of the field that are large and positive.
This can only happen if shortly after $t_2$ the field turns negative
again and does so extremely fast.  All in all we find a pattern
of ever growing oscillations.  We will now confirm this 
somewhat qualitative analysis with a computation.

\subsubsection{Rolling with ever-growing oscillations}

We now look for a solution where the tachyon rolls down from
its maximum.  We shall make the following ansatz
\be
\label{anzrolleven}
\phi(t) = 1 - \sum_{n=1}^{\infty}a_n \, e^{\alpha n t} \,,
\label{exp}
\ee
where $\alpha >0$ is a constant to be determined. Note that
we are using ``harmonics" of the basic exponential $e^{\alpha t}$.
This is sensible due to the structure of the equation we are
trying to solve. With $\alpha$ positive, moreover, we guarantee
that as $t\to -\infty$ the field is at the maximum. This kind
of expansion has also been discussed in \cite{ashokenew}.

It is possible to anticipate the value of $\alpha$. It should
correspond to $\sqrt{-M^2}$, where $M^2$ is the value of 
the mass-squared at the maximum.  This can also be derived from the
field equation, where we will confirm that $a_1$ plays the role of a
``marginal" parameter.  Since $M^2 = -2$ at the maximum (for any $p$) we
must have $\alpha =
\sqrt{2}$.  

The ansatz \refb{anzrolleven} for the solution
is plugged into the
left hand side of
(\ref{EOMBox1}) and for $p=2$ we find
\begin{equation}
\label{lsgos}
p^{{1\over2} \partial_t^2} \phi(t) =
1 -\sum_{n=1}^{\infty}a_n \, 2^{{1 \over 2} \alpha^2 n^2} \,
e^{\alpha n t}\,.
\end{equation}
In addition, the right hand side of (\ref{EOMBox1}),
letting $a_0 \equiv -1$ takes the form:
\begin{equation}
(\phi(t))^2 = \sum_{n=0}^\infty \Bigl( \sum_{m=0}^n \, a_m a_{n-m} \Bigr)
\, e^{\alpha n t}\,. 
\end{equation}
Therefore the equation gives the set of relations
\begin{equation}
\label{rels}
-a_n \, 2^{{1 \over 2} \alpha^2 n^2}= \sum_{m=0}^n \, a_m a_{n-m}\,,
\qquad  n\geq 1\,. 
\end{equation}
For the case $n=1$ we find the condition:
\begin{equation}
-a_1 2^{{1 \over 2} \alpha^2} = -2 a_1 \,,
\end{equation}
which implies that $\alpha = \sqrt{2}$, and $a_1$ is arbitrary. The
relations in \refb{rels} then give
\begin{equation}
\label{rels2}
a_n \,( 2^{ n^2}-2 ) = -\sum_{m=1}^{n-1} \, a_m a_{n-m}\, .
\qquad  n\geq 2\,. 
\end{equation}
It is clear from this equation that we can
determine the {\it exact} values of the coefficients 
$a_n$ iteratively.  For example, we readily find 
\begin{equation}
\label{a2val}
a_2 = -{a_1^2 \over 14} \,.
\end{equation}
It may seem at first sight that we have the freedom to
choose $a_1$, but a shift in time can be used to rescale it to
one (or to minus one if $a_1$ is negative, in which case the tachyon rolls
towards the unbounded side
of the potential). This
is possible because $a_n \sim a_1^n$, as can be seen from \refb{rels2}
and \refb{a2val}, and the structure of the ansatz \refb{anzrolleven}.

\begin{figure}[!ht]
\leavevmode
\begin{center}
\epsfysize=7.5cm
\epsfbox{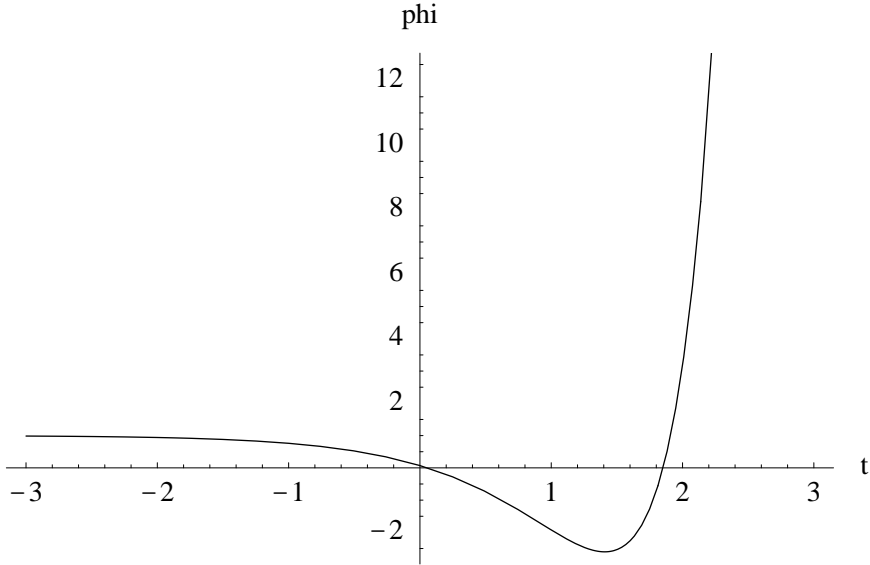}
\end{center}
\caption{The first oscillation of $\phi(t)$ as function of time. The
field oscillates with ever-growing amplitude.}
\label{expPlot}
\end{figure}

The above iterative procedure can be carried to any desired
order to find a solution that is accurate for longer and longer
times. Here we
show the result for $\phi(t)$ to level five
\begin{eqnarray}
\label{profmaxrollev}
\phi(t) &=& 1 - e^{\sqrt{2} t} + {1 \over 14} \,e^{2 \sqrt{2} t} -
{1 \over 3570} \,e^{3 \sqrt{2} t}  \nonumber\\
&&+ \,\,{283 \over 3275389320}\, e^{4
\sqrt{2} t} -{7313 \over 6105767870038200} \, 
e^{5 \sqrt{2} t} + \cdots \,\,.
\end{eqnarray}
The signs of $a_n$ alternate, that will make the field oscillate. In
figure \ref{expPlot}, we plot the first oscillation. We see that the
tachyon
starts with the value one, rolls slowly towards the minimum, then
overshoots
the point $\phi = -{1 \over 2}$ -- the point where the tachyon
would turn around in a theory with the same potential and with
a canonical kinetic term. Here the field goes down to a value
of about $\phi= -2.6$. In doing so the field has climbed a height
which is about 55 times that of the original height of the 
unstable minimum!  This can be consistent with energy conservation
because in the p-adic string model the kinetic energy can be negative.
The field turns around at $\phi = -2.6$, overshooting the maximum
and continuing to grow until it reaches a value of about
$\phi = 593$, a very large value indeed.  
At this point the potential is very negative
and therefore the kinetic energy must be very large and positive.
The field then manages to climb up the potential to go to lower
values, and  continues to oscillate with
ever-growing amplitude.

\begin{table}
\begin{center}
\begin{tabular}{|c||c|c|c|c|c|c|}  \hline
&$t_0 = 0$ & $t_0 = 1$ & $t_0 = 2$  & $t_0 = 3$  & $t_0 = 4$ &
$t_0 = 5$
\\  \hline \hline
$\phi(t_0)$ & 0.0711485 & $-1.92423$ & 3.17788 & 184.953 &
$-424.634$ & $-196021$
\\  \hline
$\phi^2(t_0)$ & 0.00506212  & 3.70267 & 10.0989 & 34207.7 & 180314 &
$3.8424 \cdot 10^{10}$ \\  \hline \hline
${\cal C}\left[\phi\right](t_0)$ & 0.00506212  & 3.70267 & 10.0989 &
34207.7 &  180314 & $3.8424 \cdot 10^{10}$
\\  \hline
\end{tabular}
\end{center}
\caption{Comparison between $\phi^2(t_0)$ and
${\cal C}\left[\phi\right](t_0)$, where $\phi$ is the solution to level
30.
The agreement is remarkable.}
\label{CheckConvTable}
\end{table}

\begin{figure}[!ht]
\leavevmode
\begin{center}
\epsfysize=7.5cm
\epsfbox{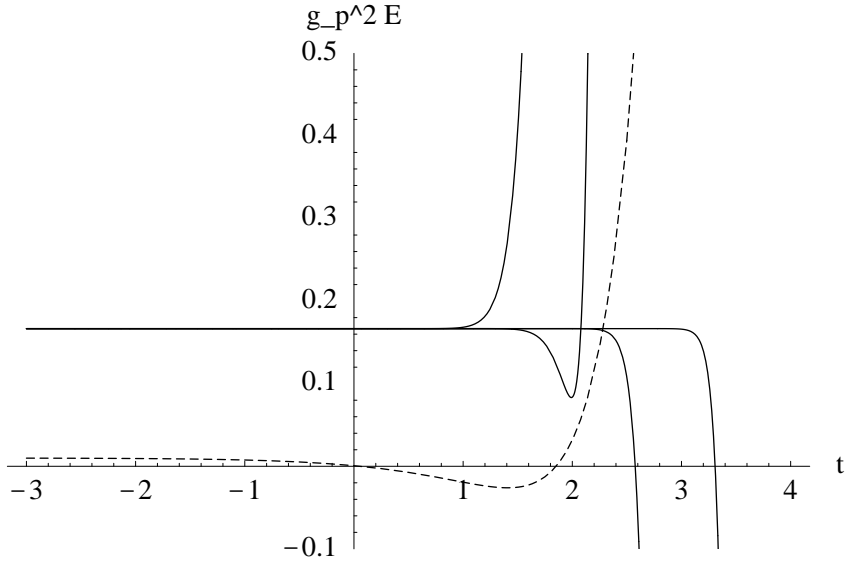}
\end{center}
\caption{Testing the constancy of the energy $E(t)$ for the
rolling solution with ever growing oscillations. This figure
shows
$g_p^2
\, E(t)$, calculated from (\ref{Epadic}),
using a level 30  solution (\ref{exp}). The successively
flatter (solid) curves are calculated from (\ref{Epadic}), keeping
respectively 40, 60, 100, and 150 time derivatives. The dashed curve is
$0.01 \phi(t)$ and is just shown as a reference.}
\label{energy}
\end{figure}

\subsubsection{Testing and exploring the solution}

We want to do two kinds of check on this solution. First, since
it was constructed from the derivative form of the equation
of motion, we want to
make sure that it satisfies the convolution form of the equation of
motion. This will also confirm the correctness of the intuition
that suggested that convolution requires ever growing oscillations.
Then we
want to measure the energy of the solution. In particular, we want
to confirm the rather unusual property that ever growing oscillations
can be compatible with the constancy of the total energy.

In table \ref{CheckConvTable}, we compare $\phi^2(t_0)$ to
the convolution ${\cal C}\left[\phi\right](t_0)$ calculated numerically.
We have taken $\phi(t)$ to level 30.
We see that the agreement holds to at least six digits accuracy.
This is impressive given that the field values include very large
numbers and therefore are sampling at least two cycles of oscillation.

\begin{figure}[!ht]
\leavevmode
\begin{center}
\epsfysize=7.5cm
\epsfbox{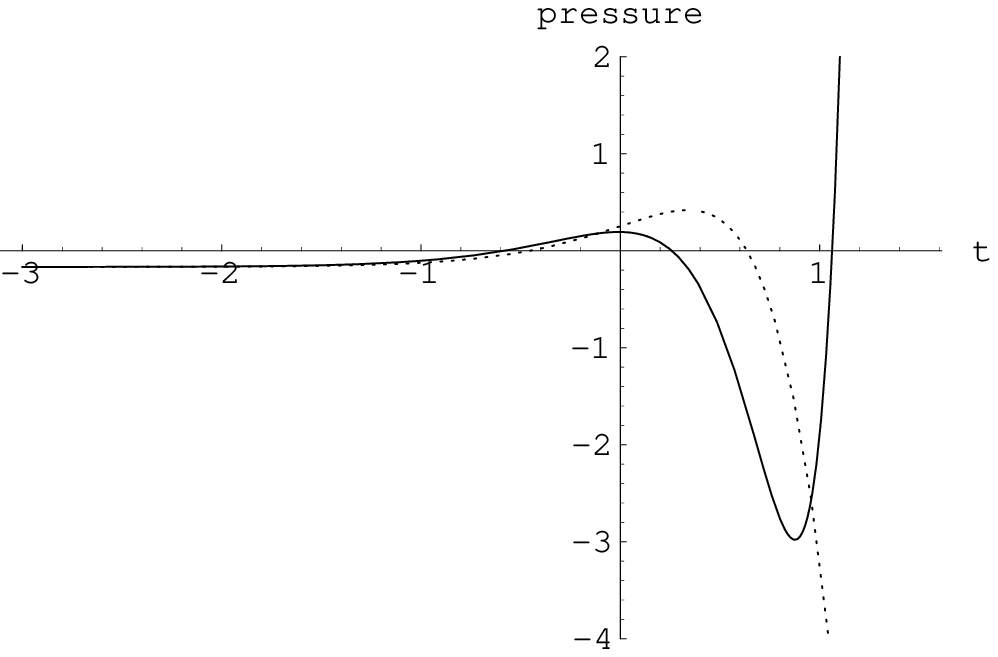}
\end{center}
\caption{Calculating the pressure of the maximal rolling solution
for the $p=2$ potential.
Shown in dashed lines is the pressure (multiplied by $g_p^2$)
as calculated 
including only first derivatives in \refb{pform}.  
In continuous line is 
the pressure including all derivatives that are relevant
in the interval shown.} 
\label{pressmaxroll}
\end{figure}

In the second test we use the energy formula
(\ref{Epadic}) to see if the solution satisfies
the requirement that the energy is constant. Since
the energy formula includes infinite number of 
derivatives we expect better accuracy as we include
more and more terms.  This is indeed what we observe
in figure \ref{energy}. 
We note that including more
derivatives
in the energy formula makes it constant for longer times (when the higher
derivatives of $\phi(t)$ become more important). It is also clear on this
graph, that the energy of the rolling solution is ${1 \over 6 \, g_p^2}$,
the height of the maximum of the potential.

In figure \ref{pressmaxroll} we show the pressure
(multiplied by $g_p^2$) as a function
of time. This was 
calculated using \refb{pform} and \refb{profmaxrollev}
evaluating the solution to level
14, and using up to 80 derivatives in the pressure formula. 
Note that for large negative
times the value is negative and equal to $(-1/6)$. Shown
in dashed lines, for comparison is the pressure including
only the first derivative contribution.  We see no sign
that the pressure is going to zero for large times, therefore
no evidence that this solution represents tachyon matter.

It is of interest to compare this solution with
similar rolling solutions in ordinary field theory.
As noted in \cite{ashokenew} the perturbative
expansion of a rolling
solution in a $\phi^3$ theory using the exponentials 
associated to the unstable maximum breaks down at a
finite time.  This does not happen with the p-adic
string solution.  The expansion coefficients $a_n$
fall off extremely fast due to the factor of 
$2^{n^2}$ in \refb{rels2}.

\sectiono{Anharmonic Oscillations Around the  Vacuum}
\label{sectAO}

The p-adic string model, as reviewed in the introduction,
satisfies one important property expected from the tachyon.
At the minimum, the mass of the field goes to infinity
and there are no conventional degrees of freedom. 
Indeed, for $\phi(t)$ small, the linearized equation
of motion is  $p^{-{1\over2} \Box} \phi = 0$, which requires
$\Box = +\infty$, that is, $m^2 = \infty$. We therefore have no
conventional harmonic oscillations at the bottom of the potential.

It will be seen, however, that while there are no oscillations
that solve the linearized equations of motion, there are
oscillatory solutions of the nonlinear equations of motion.
This is the case when $p$ is odd, and the potential is even.
Such oscillations do not appear to exist when $p$ is even.

The oscillations represent a family of solutions -- in the
limit as the frequency goes to zero they go into the kink
solution discussed earlier.
In the anharmonic oscillations to be constructed below
the amplitude of oscillation is a function of the frequency.
As the frequency goes to infinity, the amplitude will go to zero.
These oscillations can be found numerically, either by solving the
convolution equation by iteration (\ref{iteration}) starting from a
periodic
$\phi_0(t)$, or by a ``level" truncation analysis of the differential
form of the equation of motion,
applied on a Fourier expansion of $\phi$. We will explain
this last method in detail, because it will be useful in order to
understand the
analytic form of $\phi(t)$ in the large frequency limit. It will also
allow us to calculate both the amplitude/frequency relation, and the energy of the
oscillations in the large $\omega$ limit. 

\subsection{Series construction and amplitude/frequency relation}

Let us write a Fourier series for $\phi(t)$ which we imagine oscillates
with some amplitude $A$ and period $T$ around the tachyon vacuum $\phi=0$ 
of an odd $p$ potential. 
First, we choose
the origin of time axis such that $\phi(0) = A$ and $\partial_t \phi(0)=0$. Also,
since the equation of motion is invariant under $\phi \rightarrow -\phi$,
we can demand that $-\phi(t) = \phi(t+{T \over 2})$, where 
$T = {2 \pi \over \omega}$, with $\omega$ the fundamental frequency. This implies that
$\phi$ can be written in terms
of odd modes only
\be
\phi(t) = \sum_{n = 0}^{\infty}{a_{2 n + 1} \, \cos\left((2 n + 1)
\omega t\right)} \,.
\label{phicos}
\ee
Since $\omega$ is defined to be the fundamental frequency
we assume $a_1 \not=0$. It is easily seen that
\be
p^{{1 \over 2} \partial_t^2} \phi(t) = \sum_{n = 0}^{\infty}{a_{2 n + 1} \,
p^{-{1 \over 2} \omega^2 (2n+1)^2} \cos\left((2 n + 1) \omega t\right)}\,.
\label{cosder}
\ee
The right hand side $\phi^p(t)$ of the equation of motion can also be 
expressed as a sum of cosines with odd modes only because $p$ is
odd. 

\begin{figure}[!ht]
\leavevmode
\begin{center}
\epsfysize=7.5cm
\epsfbox{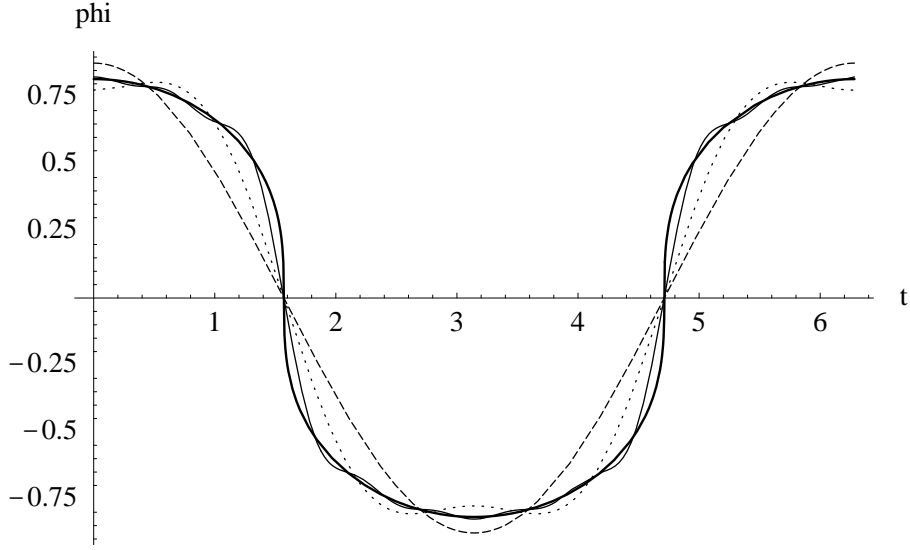}
\end{center}
\caption{Solutions for the anharmonic oscillation with $p=3$ and
$\omega=1$, to levels 1 (dashed line), 3 (dotted line), 9 (thin solid
line),
and the more accurate solution from the iteration procedure (thick line).
The
horizontal axis is time. The amplitude is $A = 0.817615$}
\label{coslevtrunc}
\end{figure}

Now we want to
truncate
$\phi(t)$ to level
$(2 N +1)$:
$\phi(t) = \sum_{n = 0}^{N}{a_{2 n + 1} \, \cos\left((2 n + 1)
\omega t\right)}$, and for concreteness we will take $p=3$.
 To level one, $\phi(t) = a_1
\cos(\omega t)$. Plugging this into the equation of motion gives
\be
p^{-{1 \over 2} \omega^2} a_1 \cos(\omega t) =
\left(a_1 \cos(\omega t) \right)^3 = {3 \over 4} \, a_1^3 \cos(\omega t)
+ {1 \over 4} \, a_1^3 \cos (3 \omega t) \,,
\ee
and, since we work to level one, we keep only the $\cos(\omega t)$ term:
\be
p^{-{1 \over 2} \omega^2} a_1 = {3 \over 4} \, a_1^3 \,.
\label{cos1}
\ee
This equation has the nontrivial solutions 
\be
a_1 = \pm {2 \over \sqrt{3}} \, 3^{-{1 \over 4} w^2} \,.
\ee
With $\omega = 1$, this is $a_1 = \pm 0.877383$.  Note that at this
level $a_1$ is the amplitude of oscillation, and it depends nontrivially
on the frequency.
To level three, $\phi(t) = a_1 \cos(\omega t) + a_3 \cos(3 \omega t)$,
and we have two equations
\ben
p^{-{\omega^2 \over 2}} a_1 &=&
{3\over4} \, a_1^3 + {3\over4} \, a_1^2 a_3 - {3\over2} a_1 a_3^2\,,
\\
p^{-{9 \omega^2 \over 2}} a_3 &=&
{1\over4} \, a_1^3 + {3\over2} \, a_1^2 a_3 - {3\over4} a_3^3 \,.
\een
With $\omega = 1$, we have, apart from 
trivial solutions and obvious sign flips,
\ben
%a_1 = 0 &,& a_3 = 0.097487 \\
a_1 = 0.930367 \,,&& a_3 = -0.153804 \,.
\een
With these values the amplitude $A \simeq 0.777$. This is not too
far from the value $A\simeq 0.818$ obtained by a rather accurate 
iterative convolution
calculation. In figure
\ref{coslevtrunc}, we show the solutions to level $1$, $3$ and $9$,
compared to the very accurate solution obtained from convolution.
At level nine the two different numerical
methods give almost the same solution. This means that the
oscillations satisfy both the convolution and the differential form of
the equations of motion. Since the $\omega
\rightarrow 0$ limit of the anharmonic oscillation is the kink, 
we also expect  that the
kink is a solution of both equations of motion, as verified
to some degree in section \ref{sectksop}.

We emphasize that
in this example we started by choosing a frequency $\omega$ and the
numerical
solution was then found to be unique (up to obvious time shifts).
In particular, the amplitude is given
uniquely in terms of $\omega$. In addition, there are solutions for
any value of the frequency. This situation is drastically different from
the case of classical harmonic oscillations in conventional field theory, where the
frequency is given by the mass of the field, but the amplitude is free to take any
value.

\begin{figure}[!ht]
\leavevmode
\begin{center}
\epsfxsize=15.0cm
\epsfbox{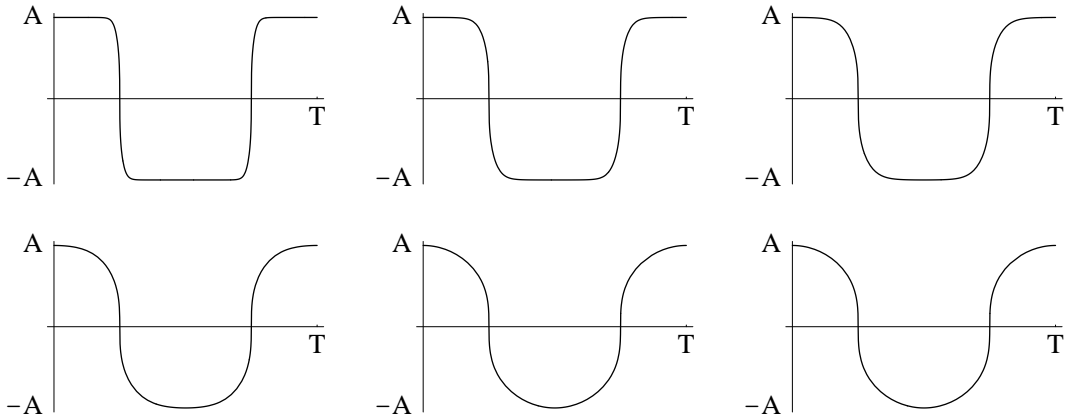}
\end{center}
\caption{One period $T$ of anharmonic oscillations for $p=3$. On the
first row from left to right: $\omega = 0.1$, $\omega = 0.2$,
and $\omega = 0.3$. On the second row from left to right:
$\omega = 0.5$, $\omega = 2$, and the last
graph shows the asymptotic profile $\left(\cos \omega t\right)^{1/3}$.
The scale has been adjusted to show all periods as equal.}
\label{oscillations}
\end{figure}

\medskip
Although we do not have a closed form solution describing the 
oscillations, we can derive such closed form in the
$\omega \rightarrow \infty$ limit.
Indeed, in this limit, the right hand side of (\ref{cosder}) is 
dominated by the first term
in the
sum: 
\be
p^{-{1 \over 2} \Box} \phi(t) = {a_1 \, p^{-{1 \over 2} \omega^2}
\cos\left(\omega t\right)} + {\cal O}\left(p^{-{9 \over 2}
\omega^2}\right)
\,.
\ee
{}From the equation of motion, we thus have
\begin{equation}
\phi^p(t) = a_1 \, p^{-{1 \over 2} \omega^2}
\cos\left(\omega t\right) + {\cal O}\left(p^{-{9 \over 2} \omega^2}\right)
\,.
\end{equation}
In the $\omega \rightarrow \infty$, the $\OO(..)$ terms can be neglected
and the above equation implies that 
\be
\phi(t) \; \stackrel{\omega \rightarrow \infty}{\longrightarrow} \;
A \left[ \cos(\omega t) \right]^{1/p} \,,
\label{profile}
\ee
where the amplitude $A$ is related to $a_1$ and to $\omega$ as:
\be
A^p = a_1 p^{-{\omega^2 \over 2}} \,.
\label{A1}
\ee
Recall that since $p$ is odd, there is no problem in
taking the root in \refb{profile} -- for negative arguments
the root is defined to be negative. The 
evaluation of 
$a_1$ is possible  from the approximate form of $\phi(t)$. Indeed $a_1$
is the  first harmonic in the expansion of  $[ \cos(\omega t) ]^{1/p}$.
Using \refb{phicos} we have
\be
A \left[ \cos(\omega t) \right]^{1/p} = 
a_1 \cos (\omega t) + a_3 \cos (3\omega t) + \cdots \,\,,
\label{phicosp}
\ee
and therefore we can calculate
\ben
a_1 = {\omega \over \pi} \int_0^{2 \pi \over \omega}{\cos(\omega t)
\phi(t) dt} &\stackrel{\omega \rightarrow \infty}{\longrightarrow}&
{A \over \pi} \int_0^{2 \pi}{\left(\cos x\right)^{1 +
{1 \over p}}  dx}
\nonumber \\
&=& A \, {2 \, \Gamma\left(1 + {1 \over 2p} \right) \over \sqrt{\pi}\,
\Gamma\left({3\over2} + {1 \over 2 p}\right)} \,.
\label{A2}
\een
Now combining (\ref{A1}) and (\ref{A2}), we find the asymptotic expression
for the amplitude:
\be
A^{p-1} \; \stackrel{\omega \rightarrow \infty}{\longrightarrow} \;
{2 \, \Gamma\left(1 + {1 \over 2p} \right) \over
\sqrt{\pi} \, \Gamma\left({3\over2} + {1 \over 2 p}\right)} \,
p^{-{\omega^2 \over 2}}
\,.
\label{amplitude}
\ee
To illustrate the asymptotic forms we show in figure \ref{oscillations}
one period of oscillations for various values of $\omega$ and  
 the asymptotic profile (\ref{profile}).

At last, in table \ref{AmpTable}, we compare the amplitudes of some $p=3$
numerical solutions (whose absolute accuracy is about $10^{-6}$) to the
asymptotic amplitudes (\ref{amplitude}).
\begin{table}
\begin{center}
\begin{tabular}{|c||c|c|c|c|c|}  \hline
&$\omega = 0.6$ & $\omega = 0.8$ & $\omega = 1$  & $\omega = 1.5$  &
$\omega = 2$ \\  \hline \hline
Numerical amplitude & 0.962253 & 0.899904 & 0.817615 & 0.580459 &
0.358948 \\  \hline \hline
Amplitude from (\ref{amplitude}) & 0.975466 & 0.903262  & 0.818225 &
0.58046 & 0.358948 \\  \hline
\end{tabular}
\end{center}
\caption{Comparison of the amplitudes of the $p=3$ anharmonic
oscillations,
calculated from the numerical solutions and from the asymptotic formula
(\ref{amplitude}).}
\label{AmpTable}
\end{table}
We see that both the profile and the amplitude approach rapidly their
asymptotic limits. The asymptotic amplitude is already precise to six
digits when $\omega \geq 2$.

\subsection{Energy of anharmonic oscillations}

It is possible to simplify the formula (\ref{Epadic}) for the energy in
the case of anharmonic oscillations. We will eventually find the
following asymptotic expression for the energy
\begin{equation}
E \; \stackrel{\omega \rightarrow \infty}{\longrightarrow} \;
{h_p \over g_p^2} \, \omega^2 \, A^{p+1} \,,
\end{equation}
where $h_p$ is a $p$-dependent constant that will be determined 
in the following calculation.

Let us plug the expansion (\ref{phicos}) into (\ref{Epadic}). Since
the energy is conserved, we can evaluate it at $t=0$; at this particular
time, all the odd derivatives of $\phi$ vanish; and the $m$-th
derivative, for $m$ even, is
\begin{equation}
\phi_m(0) = \sum_{n=0}^{\infty}{a_{2n+1} (-1)^{m\over2} (2n+1)^m
\omega^m} \,.
\end{equation}
Thus (\ref{Epadic}) becomes
\ben
E &=& -{\cal L}(t=0) - {1\over 2 g_p^2} \, \sum_{\ell = 1}^{\infty}
\left( {1\over2} \ln p \right)^\ell {1 \over \ell!}
\sum_{m=0 \atop m \, {\rm even}}^{2 \ell -2 } (-1)^m \times
\nonumber \\
&& \times \sum_{n=0}^{\infty} a_{2n+1} (-1)^{m \over 2} (2n+1)^m \omega^m
\sum_{k=0}^{\infty} a_{2k+1} (-1)^{\ell-{m \over 2}} (2k+1)^{2\ell-m}
\omega^{2\ell-m} \,.
\nonumber
\een
After obvious simplifications, and rearranging the order of summation,
we get
$$
E = -{\cal L}(t=0) - {1\over 2 g_p^2} \, \sum_{n,k=0}^{\infty}
a_{2n+1} a_{2k+1} \sum_{\ell=1}^\infty \Bigl( -{\omega^2
(2k+1)^2 \over 2} \ln p \Bigr)^\ell {1 \over \ell!}
\sum_{m=0 \atop m \, {\rm even}}^{2 \ell-2} \left( {2n+1 \over
2k+1} \right)^m \,.
$$
The last sum is a finite geometric series, it can thus be summed
explicitly
$$
\sum_{m=0 \atop m \, {\rm even}}^{2 \ell-2} \left( {2n+1 \over
2k+1} \right)^m = \left\{ \begin{array}{ll}
\ell & , \quad n=k \\
{1 - \left({2n+1 \over 2k+1}\right)^{2\ell} \over
1 - \left({2n+1 \over 2k+1}\right)^{2}} & , \quad n \neq k
\end{array} \right. \,.
$$
The energy then becomes
\ben
E &=& -{\cal L}(t=0) - {1\over 2 g_p^2} \sum_{n=0}^{\infty} a_{2n+1}^2
\Bigl(-{\omega^2 (2n+1)^2 \over 2} \ln p \Bigr)
\sum_{\ell=0}^{\infty} \Bigl(-{\omega^2 (2n+1)^2 \over 2} \ln p
\Bigr)^\ell {1 \over \ell!} -
\nonumber \\
&& - {1\over 2 g_p^2} \sum_{n,k=0 \atop n \neq k}^{\infty} a_{2n+1}
a_{2k+1} {1 \over 1-\left( {2n+1 \over 2k+1} \right)^2}
\sum_{\ell=1}^{\infty} \Bigl(-{\omega^2 \over 2} \ln p \Bigr)^\ell
{1 \over \ell!} \left( (2k+1)^{2\ell} - (2n+1)^{2\ell} \right) \,.
\nonumber
\een
We can now do explicitly the $\ell$ sums and they give simply exponentials.
And since $\phi$ is a solution of the equation of motion, we have
$$
{\cal L}(t=0) = {1\over 2 g_p^2} {1-p \over 1 + p} A^{p+1} \,,
$$
where $A = \phi(0)$ is the amplitude of the oscillation. The expression
for the energy thus simplifies to
\ben
E &=& {1\over 2 g_p^2} {p-1 \over p+1} A^{p+1} + \omega^2 \,
{\ln p \over 4 g_p^2} \sum_{n=0}^{\infty} a_{2n+1}^2 (2n+1)^2
p^{-{\omega^2 (2n+1)^2 \over 2}} +
\nonumber \\
&& + {1\over 2 g_p^2} \sum_{n,k=0 \atop n\neq k}^{\infty}
a_{2n+1} a_{2k+1} {1 \over 1-\left( {2n+1 \over 2k+1} \right)^2}
\Bigl( p^{-{\omega^2 (2n+1)^2 \over 2}} - p^{-{\omega^2 (2k+1)^2 \over 2}}
\Bigr) \,.
\label{Eoscil}
\een
We now want to find the asymptotic behavior of (\ref{Eoscil}) when
$\omega \rightarrow \infty$. This is easily done by  noting,
from (\ref{amplitude}), that $p^{-{\omega^2 \over 2}} \sim A^{p-1}$,
and from \refb{phicosp}, that $a_{2n+1}^2 \sim A^2$. With this we see
that the leading term in the first sum of (\ref{Eoscil}) is the
first term, and it is of order $\omega^2 A^{p+1}$. On the other hand 
the leading term in the second sum is of order $A^{p+1}$. This is also
the order of the first term in the equation. Therefore, in the
limit that we are considering, the leading term in the energy is
$$
{1 \over g_p^2} \, \omega^2 \, {\ln p \over 4} a_1^2
p^{-{\omega^2 \over 2}} = {1 \over g_p^2} \, {\ln p \over 2 \sqrt{\pi}}
{\Gamma\left(1 + {1 \over 2p}\right) \over
\Gamma\left({3 \over 2} + {1 \over 2p}\right)} \, \omega^2 \, A^{p+1} \,,
$$
where we have used (\ref{A2}) and (\ref{amplitude}). In total the final
expression is
\be
E = {1 \over g_p^2} \left( {\ln p \over 2 \sqrt{\pi}}
{\Gamma\left(1 + {1 \over 2p}\right) \over
\Gamma\left({3 \over 2} + {1 \over 2p}\right)} \, \omega^2 \, A^{p+1}
+{\cal O}(A^{p+1}) + {\rm higher\ powers\ of}\, A \right) \,,
\ee
where the first correction is of order $A^{p+1}$, without an
$\omega^2$ in front of it. The next terms are higher powers of $A$,
and their relative contributions are thus exponentially vanishing
when $\omega \rightarrow \infty$.  It is straightforward to confirm
that the above result would arise by computing the energy using
simply $\phi = a_1 \cos(\omega t)$. Thus in this large frequency
approximation the first harmonic carries the leading contribution
to the energy.

In harmonic oscillations, the energy goes like $A^2 \omega^2$, where
$A$ is the amplitude and $\omega$ is the frequency.  The frequency
dependence of the energy for the anharmonic oscillations is the same,
but the amplitude dependence is different -- the amplitude appears
with the same power as it appears in the potential. Of course, in 
the present case the amplitude and frequency are not independent 
variables.

In summary, we have found physical, energy-carrying excitations around
the tachyon vacuum. If we try to interpret them in the context
of string theory, they could correspond to non-conventional open
string excitations that may radiate into closed strings.
In the Sen conjecture, where the tachyon vacuum is supposed
to be the closed string vacuum, physical excitations around
the closed string vacuum would naturally be interpreted as
closed strings.  Could the above anharmonic oscillations 
represent closed strings?  Perhaps, but there are a few
complications.  First, these solutions do not have analogs
for even $p$ potentials where any oscillation must
eventually grow without limit. In some ways the even $p$ case
seems more closely related to bosonic string theory. 
In addition, we have calculated what would be the analog
of {\it classical } open string solutions. The closed string
states would correspond to the quantization of the 
above oscillations.  Since the closed string spectrum should
be coupling constant independent, the quantization must somehow
cancel out the factor $1/g_o^2$ present in the energy of the
classical solutions. For other discussion of closed strings
and the tachyon vacuum see \cite{0111129}.

\section{Family of Solutions for Even $p$}

We have already found a family of solutions for the case
of odd $p$ -- the anharmonic oscillations around the tachyon
vacuum.  In this section we will present a continuous family
of solutions for even $p$ that are, in some sense the closest
analog to the anharmonic oscillations of the odd $p$ theory.
The oscillations, however, will be unbounded. This can happen
even in the limit when the energy can be vanishingly small.

In studying the maximal rolling solution for
even $p$ in section \ref{sectMRTEP}, we saw that the ``frequency'' $\alpha$ in the
expansion (\ref{exp}) was forced to take the value $\sqrt{2}$. 
 We will now
describe a family of solutions labelled by a parameter $\alpha$ such that
$0 < \alpha < \sqrt{2}$. For simplicity we will consider only the case
$p=2$. Our ansatz is therefore 
\be
\phi(t) = \sum_{n=0}^{\infty}{a_n \cosh(n \alpha t)} \,.
\label{phicosh}
\ee
A solution of this form has time-reversal symmetry $\phi(-t) = \phi(t)$.
And at $t=0$, the tachyon field has value $\phi(0) =
\sum_{n=0}^{\infty}a_n$. Let us set up the level truncation scheme for
this ansatz. For convenience is it useful to rewrite the expansion as
\be
\phi(t) = \half \sum_{n=-\infty}^{\infty}\, \bar a_n \, e^{n \alpha t} \,,
\label{phicoshx}
\ee
with 
\begin{equation}
 \qquad
\bar a_{-n} = \bar a_n = a_n\,,\quad \hbox{for} \quad  n\not=0\,, \qquad
 \bar a_0 = 2 a_0 \,.
\end{equation}
Again, the left hand
side of the equation of motion is simple
\be
p^{{1\over2}\partial_t^2} \phi(t) =
{1\over 2} \sum_{n=-\infty}^{\infty}\,
\bar a_n\,  p^{{1\over2} n^2 \alpha^2} e^{n\alpha t}\,.
\ee
The right hand side of the equation of motion gives:
\be
\phi^2(t) = {1\over 4} \sum_{n= -\infty}^\infty \Bigl( \sum_{m=-\infty}^\infty
\bar a_m \, \bar a_{n-m} \Bigr) \, e^{n\alpha t} \,.
\ee
The last two equations imply that the equation of motion is equivalent
to the set of equations:
\be
\bar a_n\,  p^{{1\over2} n^2 \alpha^2} = {1\over 2} 
\sum_{m=-\infty}^\infty
\bar a_m \, \bar a_{n-m} \,.
\ee
The structure of these equations ensures that it suffices
to consider $n\geq 0$.  Moreover it is clear that given
that the sum in the right hand side is infinite one cannot
solve these equations exactly. A level expansion, of course, 
is possible. To work at level $N$ means to keep all equations
above with $0\leq n\leq N$ and to keep all $\bar a_q$ with
$|q| \leq N$.  Finally, let us note that
given a solution with expansion coefficients $a_n$ we get another solution
by letting 
\be
\label{signflip}
a_n \to  (-1)^n a_n \,.
\ee

To level zero the
equation of motion is
$a_0 = a_0^2$, and has for solution a tachyon sitting either at the maximum or at the
minimum of the potential. To level one, we have two equations for
$a_0$ and $a_1$
\ben
&& a_0 = a_0^2 + {a_1^2 \over 2} \,, \\
&& 2^{{1\over2} \alpha^2} a_1 = 2 a_0 a_1 \,.
\een
Apart from trivial solutions, one finds
\ben
a_0 &=& 2^{{\alpha^2 \over 2}-1} \,,
\nonumber \\
a_1 &=& \pm 2^{{\alpha^2 \over 4} - {1\over2}}
\, \sqrt{2 - 2^{\alpha^2 \over 2}}  \,.
\label{coshlev1}
\een
Some interesting features can already be seen to this level. First of all,
note that we have two branches, indexed by the sign of $a_1$, and
as could be anticipated by \refb{signflip}. If $a_1$ is
positive, a higher level analysis indicates 
that so will be all higher coefficients $a_n$,
and the solution diverges to positive field values without oscillating. 
When $a_1$ is negative, the signs of the $a_n$ will alternate, 
making the solution oscillate
with ever-increasing amplitude.

It is also clear from (\ref{coshlev1})  that $\alpha$ has to be
smaller than $\sqrt{2}$ in order for $a_1$ to be real. In this regard,
the maximal rolling solution of section \ref{sectMRTEP}, having 
$\alpha = \sqrt{2}$
can be thought as a limit case. It cannot be obtained, however, by taking
the $\alpha \rightarrow \sqrt{2}$ limit of (\ref{phicosh}), that would
only give a trivial solution.

\begin{figure}[!ht]
\leavevmode
\begin{center}
\epsfysize=7.5cm
\epsfbox{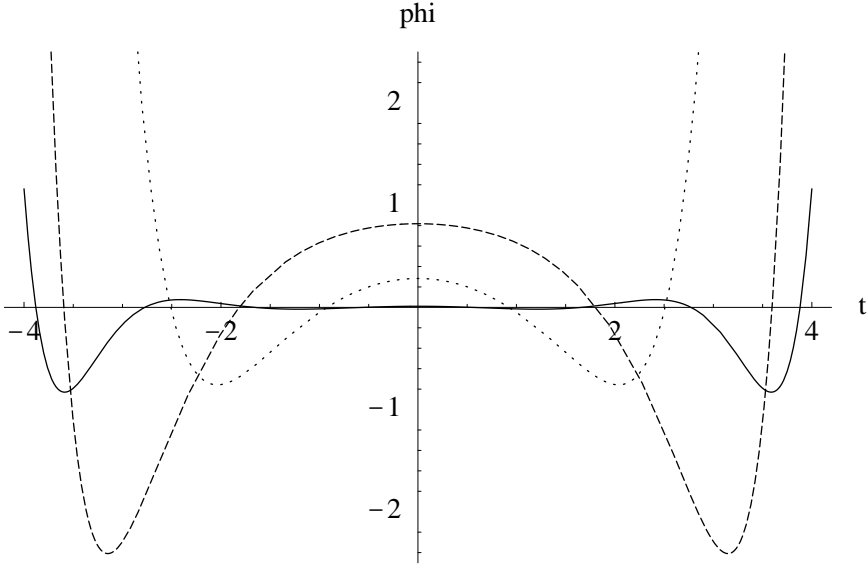}
\end{center}
\caption{Tachyon profiles as functions of $t$ for $\alpha = 1.4$
(dashed line), $\alpha = 1.2$ (dotted line), and $\alpha = 0.8$ (solid
line).
The curves become flatter near the origin when $\alpha$ is decreased.}
\label{coshes}
\end{figure}

The values of the field at $t=0$ and at this present level one approximation are
(taking $a_1$ to be negative)
\be
\lim_{\alpha \rightarrow \sqrt{2}} \phi(0) = 1\,, \qquad
\lim_{\alpha \rightarrow 0} \phi(0) = {1\over2} - {1\over \sqrt{2}}
\approx -0.207 \,.
\ee
Higher level analysis shows that for any $\alpha$ between $0$ and
$\sqrt{2}$, the solution oscillates without a bound. When
$\alpha \rightarrow \sqrt{2}$, we have $\phi(0) \rightarrow 1$, and
the energy tends to ${1 \over 6 \, g_p^2}$, the height of the potential
at its
maximum. When $\alpha \rightarrow 0$ however, $\phi(0) \rightarrow 0$
and the field becomes flatter near $t=0$, in other words one has to
wait for longer times before we see it taking large values. This is
illustrated in figure \ref{coshes}. Moreover the energy goes to zero 
very fast as $\alpha$ becomes small. Even though in these solutions
the field can spend a long time oscillating around the tachyon vacuum
their energy is necessarily very small, and thus they cannot represent
tachyon matter.  In
table \ref{EcoshTable}, we show a few values calculated from the formula
(\ref{Epadic}) applied on numerical solutions at level 30.

\begin{table}
\begin{center}
\begin{tabular}{|c||c|c|c|c|c|c|}  \hline
$\alpha$ & 1.41 & 1.4 & 1.3  & 1.2 & 1 & 0.8
\\  \hline \hline
$g_p^2 \, E$ & 0.16082 & 0.147657 & 0.0592576 & 0.0205954 & 0.00114305 &
$6.59996 \cdot 10^{-6}$ \\  \hline
\end{tabular}
\end{center}
\caption{Energy of the solution for some values of the parameter $\alpha$.
When $\alpha$ is close to $\sqrt{2}$, we have $g_p^2 \, E \approx
{1\over 6}$.
And as $\alpha$ decreases, the energy goes to zero very fast.}
\label{EcoshTable}
\end{table}

\sectiono{Rolling the Tachyon in Open String Field Theory}
\label{sectOSFT}

In this section we study the rolling of the tachyon
in OSFT. There are at least two ways we could do this 
analysis.  In the first one, we would use level expansion
to focus on a few low levels, write the equations of 
motion (containing infinitely many time derivatives) and
solve them as in the p-adic string case.  A second method
was proposed in ref.~\cite{Sen:2002nu} (section 2), and it uses   
the analytic continuation of the marginal string field
theory solution of ref.\cite{0007153}. This is the method
we will follow here to examine the rolling tachyon. 

We will consider the solution representing
a marginal deformation of a D1 brane stretched along a circle 
of unit radius. This marginal deformation
can be used to interpolate
from the D1 into a D0 brane, and it involves tachyon 
and higher field harmonics of the type $\cos (nX)$, where
$X$ is a coordinate along the circle.  The analytic continuation
requires replacing $\cos (nX)\to \cosh (nt)$. We use the solutions
of \cite{0007153}, section 3, and let the radius go to one. In addition,
we work including fields up to level four and  interactions up to
level eight, using the action given in Appendix B of \cite{0007153}. In this
way we can include up to the second harmonic of the tachyon. The marginal
direction is represented by the first tachyon harmonic
$T_1$ appearing in the tachyon field expansion
\begin{equation}
T(t) = T_0 + T_1 \cosh (t)  + T_2 \cosh(2t) + \cdots \,.
\end{equation}

\begin{figure}[!ht]
\leavevmode
\begin{center}
\epsfysize=7.5cm
\epsfbox{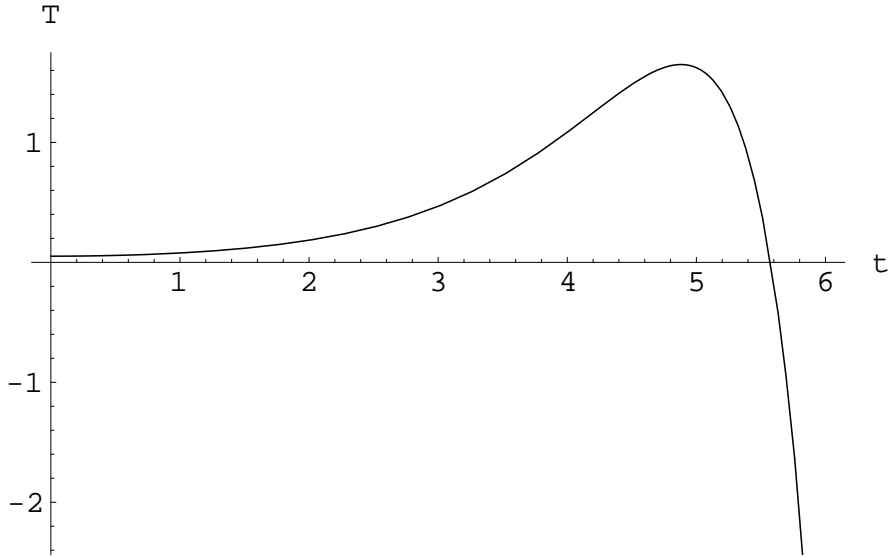}
\end{center}
\caption{The tachyon profile as a function of time as it
rolls down in OSFT. The initial conditions are 
$T(0)=0.05144$ and $T'(0)=0$.  The tachyon rolls past the
tachyon vacuum $T\simeq 0.55$ and the first turning point
is at about $T \simeq 1.65$.}
\label{presstime}
\end{figure}

The string field equations are solved by setting a value for
$T_1$ and solving for the other components of the string field.
We take $T_1 = 0.05$ and the equations of motion give
\begin{equation}
\label{tevol}
T(t) = 0.00162997 + 0.05\cosh (t)  - 0.000189714 \cosh(2t) \,.
\end{equation}
It is in fact this result that motivated much of the analysis
in the present paper, and a few comments are thus in order. 
First note that use of the $\cosh$ expansion implies that the
field satisfies $\dot T (0) =0$. The other initial condition could
be viewed as giving the value of the tachyon at $t=0$. But certainly
that is not the way the equation is solved. Fixing $T_1$ one solves
for all other harmonics $T_0, T_2, \cdots $, and all of them together
fix the initial value of the tachyon.  For $T_1=0.05$ we find from
the above equation that $T(0)=0.05144$ and $T'(0)=0$.  The first harmonic 
$\cosh (t)$ reflects the blowing up of the tachyon in the approximation
where the potential is quadratic. Note that the second harmonic
comes about with negative sign, thus at some time the tachyon
stops growing and turns around.  We had expected naively that
this turning point would be around the tachyon vacuum, thus signaling
in this approximation that the tachyon does not cross over. But
the result does not support this expectation. In
here the tachyon overshoots the tachyon vacuum by a large factor.  The
maximum is reached for
$t\simeq 4.88$ and gives $T= 1.649$, which is about three times the value 
$T \simeq 0.55$ representing the tachyon vacuum. The tachyon as
a function of time, is shown in figure \ref{presstime}. 
We have verified that the turning
point $T\simeq 1.65$ is quite stable under 
small changes of the marginal parameter $T_1$,
supporting the conclusion that the tachyon generically overshoots the
tachyon vacuum before turning around.  

The analogy with the $p=2$ maximal rolling solution is 
actually quite striking. 
Recall the discussion below eqn.~\refb{profmaxrollev}, where we observed
that the p-adic tachyon climbs a height equal to about
55 times the height of the unstable vacuum.  We can estimate
easily the corresponding  quantity in the above OSFT computation.
Taking the tachyon potential to be the level zero one
$V = -{1\over 2} T^2 + {27\sqrt{3}\over 64} T^3$, an overshooting
to $T= 1.65$ implies that the OSFT tachyon climbs up about 56
times the height of the unstable vacuum! While such close agreement
is most likely a coincidence, the fact that these quantities
are close confirms that the p-adic model seems to capture
quite well OSFT features in a simpler context. 

What happens next?  To answer
this we would need the value of $T_3$ the third tachyon harmonic.
This, however, would require a level (9,18) computation which is a
major new task. There are indications, however, that the next coefficient in
\refb{tevol} would be positive, and that ever growing oscillations
occur, just as in the case of p-adic strings with even $p$.

For the value $T_1 =0.05$ used above, the next three fields
$u,v$ and $w$, all of them at level 2, are:
\begin{eqnarray}
\label{uvw}
u (t) &=& 0.000663831 + 9.216\times 10^{-7} \cosh (t) \,, \nonumber \\
v(t) &=& -0.00162516 + 0.00003999\cosh (t) \,, \\
w(t) &=& 0.000301312 -6.325 \times 10^{-6} \cosh (t) \,. \nonumber
\end{eqnarray} 
With such small values for the first harmonic, these fields
do not appear to change significantly the physics of the tachyon.
This was also the case in the original marginal deformation problem,
where level expansion was seen to operate quite well.
In this level (4,8)
solution there are additional fields at level three and at level four. 
We will not write here their values.

We will not attempt here a computation of the pressure
in this OSFT solution, but such computation would be of
interest. With the behavior of the solution strikingly
similar to that of the  p-adic string solution for $p=2$,
we see no reason to expect that the pressure goes to zero 
asymptotically, but there could be surprises.
A computation would settle  this important issue.

\sectiono{Concluding Remarks and Open Questions}

In this paper we have studied tachyon dynamics in
string field theory and in p-adic string theory
facing up to the novelties due to the infinite number
of time derivatives in the field equations.

Several classes of rolling solutions of p-adic
string theory have been studied. We have not been
exhaustive in showing all of them nor in trying
to show that we have a complete set of solutions.
For example, there are non-oscillatory diverging
solutions for $p=2$ related by the sign change in
\refb{signflip} to the oscillatory solutions.  
In addition there are similar solutions for $p=3$
(and possibly all odd $p$). These solutions 
also show ever growing oscillations, and can
be modified by the sign change in \refb{signflip}
and by an overall change of sign. There are also
additional solutions (both for even and odd $p$)
arising from the analytic continuation of the
familiar lump solutions of p-adic string theory.
At present, for any value of the energy we know of
a finite number of solutions.

In general terms our analysis has revealed a few
surprises:

\begin{itemize}

\item  The solution space of the  equations of motions
appears to consist of real analytic (smooth) functions.
Solutions  appear to be in
one-to-one correspondence with {\it consistent} initial values for
the field and all of its derivatives.
The dynamical equation imposes a possibly infinite number of conditions
on the initial values and thus restricts  considerably the
solution space.  This remarkable feature
of the equation of motion  was
not anticipated by the consideration of systems with finite
number of time derivatives.

\item  There are energy-carrying solutions of the
field equations around the tachyon vacuum.  These 
bounded but anharmonic oscillations shown to exist
in the case of p-adic strings with even potentials,
represent solutions of the non-linear equations not 
anticipated by the
linearized (cohomology-like) problem that admits no
solutions.  

\item  There exist oscillatory solutions with ever-growing
amplitudes and constant energy. In fact, such solutions
exist even for energies that approach zero. These solutions
exhibit behaviors not seen in  lagrangians quadratic in
first time derivatives where the  kinetic energy is positive definite.
In the systems considered here the kinetic energy can be negative
and thus one can see the tachyon move to higher and higher
heights on the tachyon potential while conserving the total 
energy.
 
\end{itemize}

At the technical level we have calculated
the energy-momentum tensor in higher derivative theories,
finding useful expressions for the energy  and the pressure
in the case of time-dependent solutions. We have also
learned how to extract qualitative behavior of the solutions
using the convolution form of the field equation.
Finally, we have shown how to obtain solutions by numerical 
methods, and by analytic methods in certain limits.  

Interesting open questions remain. We list below a few of
them:

\begin{itemize}

\item
From the physical viewpoint, the most puzzling result
has been that none of the rolling solutions obtained
here appear to represent tachyon matter. That is, we have
not found solutions where with varying values of the
energy the pressure goes to zero for long times.  The rolling solutions
described in this paper would seem to yield oscillating
energy-momentum tensors that could effectively convert
the D-brane energy into closed string excitations. This had
been the conventional wisdom (see, for example \cite{Gutperle:2002ai}) 
before studies of tachyon matter.
Given the exciting possibility that a stable form of tachyon matter could
have astrophysical consequences it seems of utmost importance
to confirm its (theoretical!) existence using string field theory. 
Since the tachyon vacuum is at a finite and apparently regular
value in the field space of string field theory, 
tachyon matter must be, if present,
an exotic solution.

\item
For the case of even p-adic potentials, the anharmonic 
excitations around the tachyon vacuum do represent
non-conventional type of excitations. The most straightforward
interpretation is that they are degrees of freedom
that would radiate their energy into closed strings.
Could they actually represent closed strings? While preliminary
indications do not support this interpretation, further investigation
of the physics and interpretation of these solutions
are of interest.

\item 
We have seen that there is no conventional initial
value problem in the sense that finding {\it consistent 
initial conditions} appears to be the same problem
as finding the solution. The space of consistent
initial conditions appears to be strongly constrained
in the nonlinear equations of motion of p-adic string
theory.
How many parameters does the space of consistent initial
conditions have?

\item Our work has given an explicit construction
for the energy density of spatially homogeneous solutions
in OSFT. The calculation of the pressure was more subtle
and was only done for the p-adic string.
The OSFT pressure should be computed to test if, as 
opposed to the case of the p-adic model, the ever growing
oscillations could have an asymptotically vanishing pressure. 
A more geometrical construction 
would use open/closed string field theory \cite{9705241}, where
the  energy-momentum tensor would arise from  a variation of the 
closed string field. It may be of interest to give 
such stringy construction.

\item  One feature of string field theory is that
the infinitely many time derivatives of the covariant
formulation turn into first order time derivatives
in the light cone formulation. Studies of gauge 
fixing in string field theory have not revealed 
how this transformation actually happens.  It would therefore
be of interest to understand if there is 
a light-cone gauge p-adic string theory. If the answer
is affirmative the passage from covariant to light-cone
formulation could be tractable and provide much insight.

\item  The convolution form of the field equation 
seems to hide completely the causality properties 
of the time evolution. It would be of interest to
formulate and explore a test of causality in p-adic string theory.

\end{itemize} 

It is clear that in p-adic string theory and
in OSFT we have a rich set of problems and 
issues that should allow us to improve our
understanding of  time evolution
in string theory. Such understanding seems essential
for the development of string theory cosmology.

\bigskip
\noindent
{\bf Acknowledgements.}
We would like to thank Joe Minahan and Ashoke Sen for many
stimulating conversations and discussions throughout this work and
critical comments on a draft. 
N.~M. wants to thank Dru
Renner for help with the C language.  B.~Z. would like to acknowledge
helpful conversations with M.~Headrick, E.~Martinec and N.~Nekrasov
at Cargese 2002.
This work  was supported in part
by DOE contract \#DE-FC02-94ER40818.

\end{document}